\documentclass[twocolumn,showpacs,pra,amsmath,amssymb,floatfix]{revtex4-1}
\usepackage{amsmath,amssymb,color}
\usepackage{bm}
\usepackage{epsfig}
\usepackage{psfrag}
\usepackage{verbatim}
\usepackage{lipsum}
\usepackage{mhchem}

\newcommand{\bd}{\bm}

\newcommand{\rFactor}{\gamma}

\begin{document}

\title{Second-order interaction corrections to the Fermi surface and the quasiparticle
properties of dipolar fermions in three dimensions}

\author{Jan Krieg}
\affiliation{Institut f\"{u}r Theoretische Physik, Universit\"{a}t
  Frankfurt,  Max-von-Laue Strasse 1, 60438 Frankfurt, Germany}

\author{Philipp Lange}
\affiliation{Institut f\"{u}r Theoretische Physik, Universit\"{a}t
  Frankfurt,  Max-von-Laue Strasse 1, 60438 Frankfurt, Germany}
  
\author{Lorenz Bartosch}
\affiliation{Institut f\"{u}r Theoretische Physik, Universit\"{a}t
  Frankfurt,  Max-von-Laue Strasse 1, 60438 Frankfurt, Germany}

\author{Peter Kopietz}
\affiliation{Institut f\"{u}r Theoretische Physik, Universit\"{a}t
  Frankfurt,  Max-von-Laue Strasse 1, 60438 Frankfurt, Germany}

\date{March 3, 2015}

 \begin{abstract}

We calculate the renormalized Fermi surface and
the  quasiparticle properties in the Fermi liquid phase of  
three-dimensional dipolar fermions
to second order in the dipole-dipole interaction. 
Using parameters relevant to an ultracold gas of
erbium atoms, we find that the second-order corrections 
typically renormalize the
Hartree-Fock  results by less than one percent. 
On the other hand, if we use the second-order correction  to the compressibility
to estimate the regime of stability of the system, the point of instability
is already reached for a significantly 
smaller interaction strength than in the Hartree-Fock approximation.

\end{abstract}


\maketitle

\section{Introduction}
\label{subsec:introduction}

During the last decade, the field of ultracold fermionic gases with large dipole-dipole interaction has seen rapid advances. Atomic dipolar gases have been created using \ce{^{53}Cr}~\cite{chicireanu2006simultaneous}, \ce{^{161}Dy}~\cite{Lu2012,Burdick2014}, and \ce{^{167}Er}~\cite{Aikawa2014,aikawa2014observation}, while molecular dipolar gases have been realized with RbCs~\cite{Sage2005}, LiCs~\cite{Deiglmayr2008,Deiglmayr2010,Repp2013}, NaLi~\cite{Heo2012}, NaK~\cite{Wu2012}, and KRb~\cite{Ospelkaus2008,Ni2008,ade2009ultracold,ni2010dipolar,DeMiranda2011}. This has caused a surge of theoretical interest in these systems~\cite{Bar08,Bar12,Cha10,Yamaguchi2010,Lu2013,Goral2001,Miy08,Fregoso2009,Fregoso2010,Ronen10,Baillie2010}.
Calculations in two~\cite{Cha10,Yamaguchi2010,Lu2013} and three~\cite{Goral2001,Miy08,Fregoso2009,Fregoso2010,Cha10,Ronen10,Baillie2010} dimensions have led to the prediction that in the regime
where the Fermi liquid phase is stable,
the anisotropy of the dipolar interaction leads to a nematic deformation 
of the Fermi surface as well as to anisotropic quasiparticle properties.
Very recently  this prediction has partly been 
confirmed experimentally for a three-dimensional system 
by Aikawa {\it{et al.}}~\cite{aikawa2014observation}, who 
cooled fermionic $^{167}\text{Er}$ atoms, confined in a three-dimensional harmonic trap, well below the 
Fermi temperature $T_F$ and probed them via time-of-flight measurements. 
They found that the Fermi surface indeed elongates along the direction of the external field, in good agreement with theoretical predictions.

Due to the partly attractive nature of the dipole-dipole interaction, it is expected that for a strong enough interaction (or high enough density) the normal Fermi liquid phase becomes unstable, giving rise to
superfluid~\cite{Baranov2002a,Bruun2008,Cooper2009,Zhao2010,Liao2010}, liquid crystalline~\cite{Fregoso2009,Fregoso2009a,Quintanilla2009,Carr2009,Maeda2013}, density wave~\cite{Sun2010,Mikelsons2011,Parish2012,Block2014}, or Wigner crystal phases~\cite{Matveeva2012,Babadi2013}.
In contrast, for a rapidly rotating 2D system of dipolar fermions the Wigner crystal phase is possibly the ground state for low densities, while for higher densities it turns into a Laughlin liquid state~\cite{Baranov2005,baranov2008wigner}. Further studies focused on finite-temperature effects~\cite{zhang2010thermodynamic,Endo2010,Baillie2010,Kestner2010,Zhang2011}, dynamical properties in the collisionless and hydrodynamic regimes~\cite{Goral2003,Sogo2009,Lima2010a,Lima2010,Abad2012,Babadi2012,Wachtler2013}, bilayer configurations~\cite{Pikovski2010,baranov2011bilayer,Zinner2012,Matveeva2014}, and quench dynamics~\cite{Nessi2014}. However, less attention has been paid to the 
quasiparticle properties in the Fermi liquid phase beyond the mean-field level. 
While this has been studied for isotropic two-dimensional
systems~\cite{Matveeva2012,Lu2012a}, to our knowledge no comparable work exists in three dimensions. Liu and Yin~\cite{Liu2011a} computed an approximation for the correlation energy and the resulting corrections to the stability limit of the system; however, they did not obtain corrections to the quasiparticle properties.

This has motivated us
to calculate the self-energy $\Sigma(\bd{k}, \omega)$ 
of three-dimensional dipolar fermions
to second order in the interaction, which is the lowest order  where 
the self-energy acquires a frequency dependence,
leading to a reduced quasiparticle weight and
a finite lifetime of the quasiparticles. But also the  shape of the Fermi surface and the
renormalized Fermi velocity receive second-order corrections which are not taken into account
in a self-consistent Hartree-Fock approximation. The purpose of this work
is to give a quantitative estimate of the size of these second-order effects.
Moreover, we shall also calculate the  renormalized chemical potential as a function of the density
to second order in the interaction, which allows us to estimate the compressibility and thus the
interaction strength where the normal Fermi liquid phase of the dipolar many-body system becomes unstable in the density-density channel.

\section{First-order self-energy}
\label{sec:firstorder}

\begin{figure}
\includegraphics[width=\linewidth]{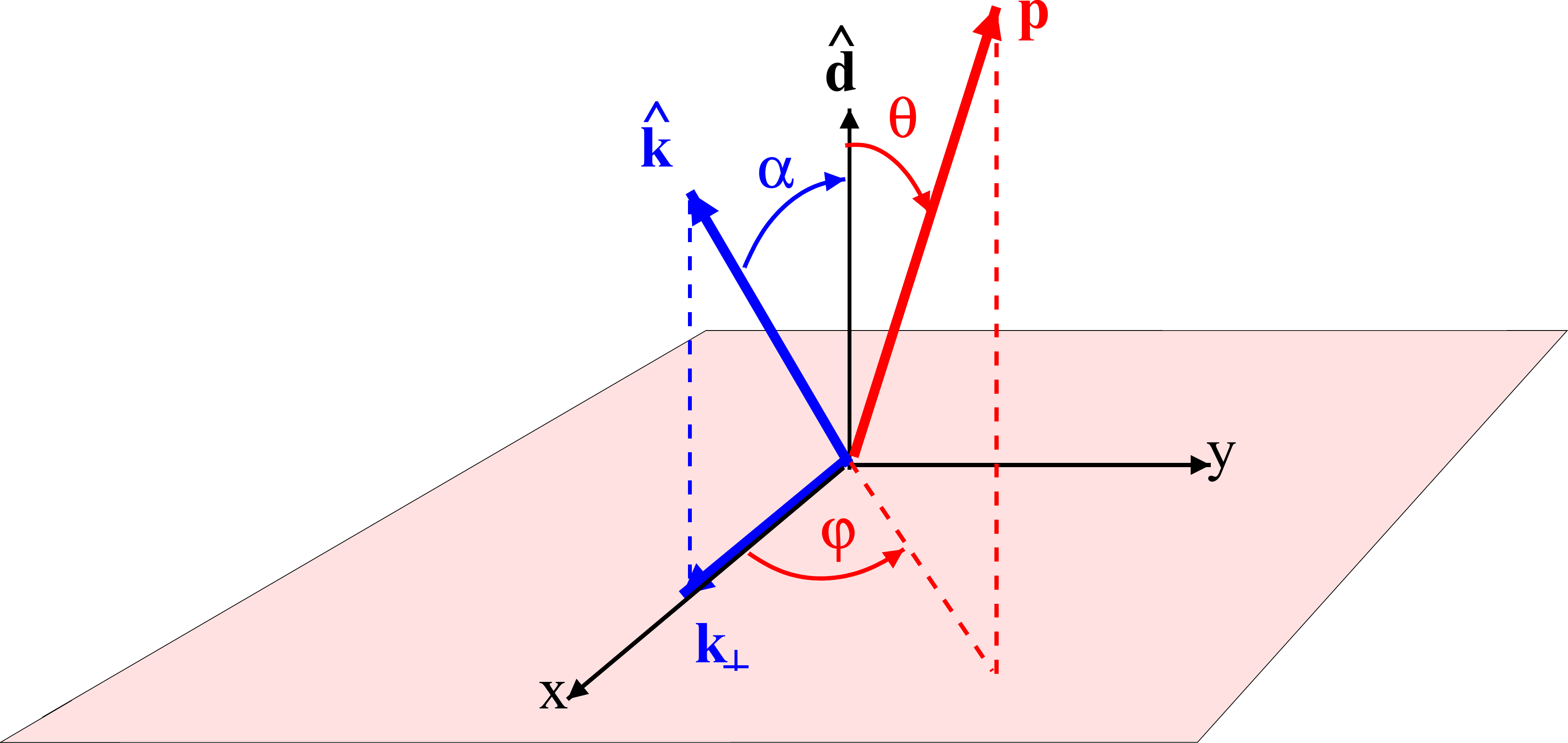}
\caption{
(Color online)
For the explicit evaluation of the self-energy
we choose our coordinate system such that the $z$ axis points in the direction $\hat{\bd{d}}$ of the dipoles and the $x$ axis lies in the plane spanned by $\hat{\bd{d}}$ and $\hat{\bd{k}}$. We call $\hat{\bd{d}} \cdot \hat{\bd{k}} = \cos \alpha$, ${\bd{k}}_{\bot} = \sin \alpha \hat{\bd{x}}$, and parametrize  the integration vector $\bd{p}$ in Eq.~(\ref{eq:I3}) as $\bd{p} = p [ \cos \theta  \hat{\bd{d}} + \sin \theta ( \cos \varphi \hat{\bd{x}} + \sin \varphi \hat{\bd{y}}) ]$.
}
\label{fig:dipole3}
\end{figure}

Before embarking on the calculation of the second-order self-energy,
it is instructive to review the evaluation of the self-energy to first order in the interaction \cite{Cha10,Fregoso2010}.
We consider a system of single-component fermions
which interact via dipolar forces in three dimensions and assume that the dipole moments $\bd{d} = d \hat{\bd{d}}$ are aligned by an external magnetic or electric field in direction $\hat{\bd{d}}$. The system is then described  by the second-quantized Hamiltonian
\begin{eqnarray}
 {{\cal{H}}} & = &   \int d^3 r \, \hat{\psi}^{\dagger} ( {\bd{r}} )
  \left( - \frac{\nabla^2}{2m} \right) \hat{\psi} ( {\bd{r}} ) 
 \nonumber
 \\*
 & + &
 \frac{1}{2} \int d^3 r \int d^3 r^{\prime} \,
\hat{\psi}^{\dagger} ( {\bd{r}} )\hat{\psi} ( {\bd{r}} )
 U ( {\bd{r}} - {\bd{r}}^{\prime} )  \hat{\psi}^{\dagger} ( {\bd{r}}^{\prime} )   
\hat{\psi} ( {\bd{r}}^{\prime} ),
 \nonumber
 \\*
 & &
 \label{eq:hamiltonian}
\end{eqnarray}
where $\hat{\psi} ( {\bd{r}} )$ annihilates a fermion at position ${\bd{r}}$ and
we set $\hbar = 1$ throughout the paper. The dipole-dipole interaction is given by
 \begin{equation}
 U ( {\bd{r}} ) =   \frac{d^2}{| \bd{r}  |^3}  \left[ 1   - 3 ( \hat{\bd{d}} \cdot \hat{\bd{r}} )^2 \right] = - \frac{2 d^2}{| \bd{r}  |^3} P_2 ( \hat{\bd{d}} \cdot \hat{\bd{r}} ),
 \label{eq:dip}
 \end{equation}
with $\hat{\bd{r}} = \bd{r} / | \bd{r} |$ and the second Legendre polynomial $P_2 (x) = (3 x^2 - 1) / 2$.
Assuming that the system is 
confined to a box with volume $V$ with periodic boundary conditions, it is convenient to 
expand the field operators in plane waves,
$ \hat{\psi} ( \bd{r} ) =  \frac{1}{\sqrt{V}} \sum_{\bd{k}}  e^{ i \bd{k} \cdot \bd{r} }
 {c}_{\bd{k}}.$
Then the Hamiltonian (\ref{eq:hamiltonian}) can be written in momentum space as follows,
  \begin{eqnarray}
 {{\cal{H}}} 
 & = & \sum_{\bd{k}} \epsilon_{\bd{k}} {c}^{\dagger}_{\bd{k}} {c}_{\bd{k}}
 + \frac{1}{2V} \sum_{  \bd{q} } U_{\bd{q}}  {\rho}_{- \bd{q}} {\rho}_{\bd{q}},
 \label{eq:Hk}
 \end{eqnarray}
where $\epsilon_{\bd{k}} = \bd{k}^2 /(2m)$ is the free fermion dispersion,
the operators
$ {\rho}_{\bd{q}} = \sum_{\bd{k}}  {c}^{\dagger}_{\bd{k}} {c}_{\bd{k} + \bd{q}}$
represent the Fourier components of the density, and
 \begin{equation}
 \label{eq:fourier_transform_3d}
 U_{\bd{q}} = \int d^3 r e^{ - i  \bd{q} \cdot \bd{r}} U ( \bd{r} ) =\frac{8 \pi d^2 }{3} P_2 ( \hat{\bd{d}} \cdot \hat{\bd{q}} )
 \end{equation}
is the Fourier transform of the interaction. Here and below the unit vectors are denoted by $\hat{\bd{q}} = \bd{q} / | \bd{q} |$. Due to the explicit breaking of the rotational invariance by the dipolar interaction, the self-energy $\Sigma ( \bd{k} , \omega )$ not only depends on the absolute value $| \bd{k} |$ of the momentum $\bd{k}$, but also on the angle $\alpha$ between the two vectors $ \bd{d}$ and $\bd{k}$ shown in  Fig.~\ref{fig:dipole3}.

\begin{figure}
\includegraphics[width=\linewidth]{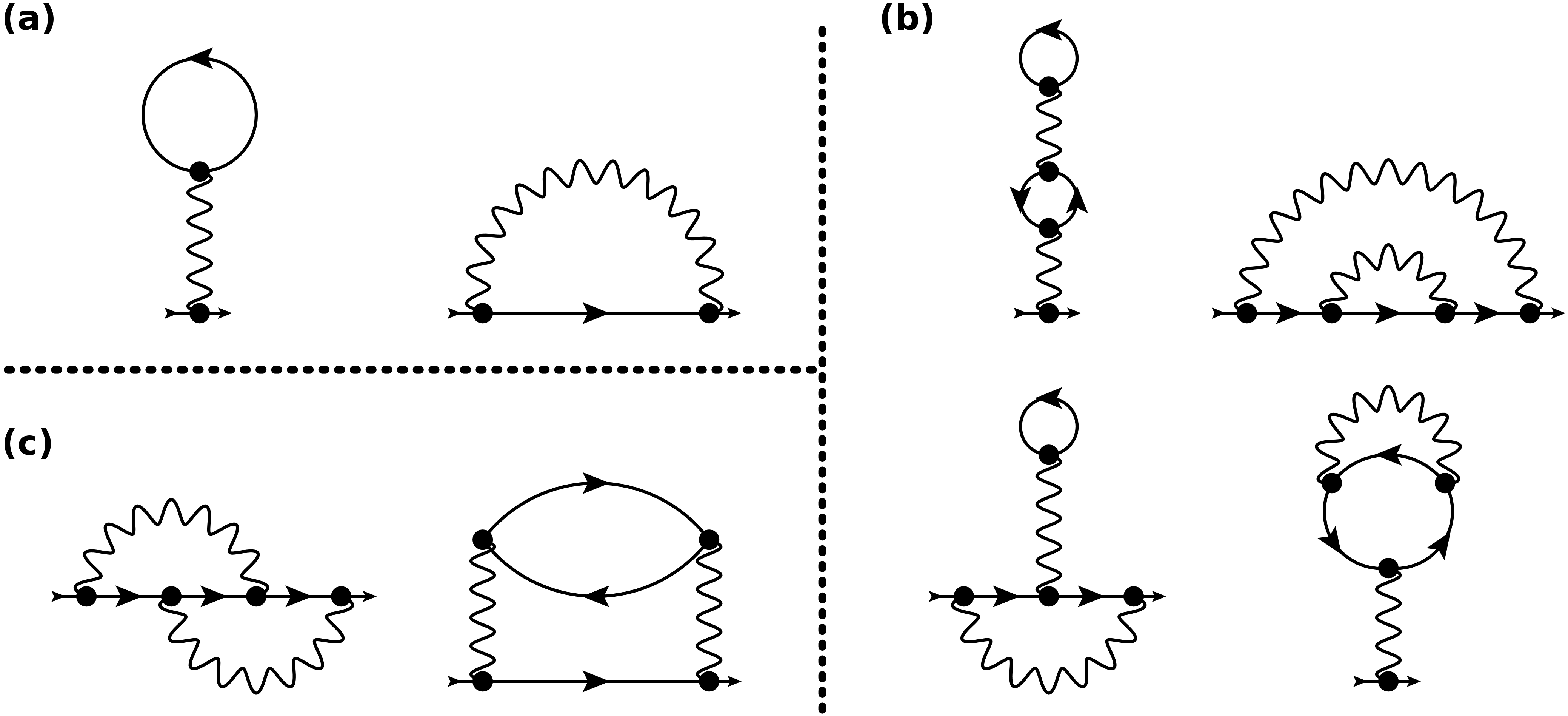}
\caption{
Relevant Feynman diagrams: (a) first-order Hartree-Fock diagrams; (b) second-order diagrams 
generated by the self-consistent Hartree-Fock approximation; (c) additional (frequency dependent) 
second-order diagrams.
Here solid lines denote the bare propagator while wavy lines denote the dipole-dipole interaction.
}
\label{fig:diagrams}
\end{figure}

To first order in the interaction, the irreducible self-energy is given by
 \begin{equation}
 \Sigma^{(1)} ( \bd{k} ) = \int \frac{ d^3 q}{(2 \pi)^3}
 ( U_0 - U_{ \bd{q}} ) f ( \epsilon_{\bd{k} + \bd{q}} ),
 \label{eq:sigma1}
 \end{equation}
where
  $
 f  ( \epsilon ) = [ e^{\beta ( \epsilon - \mu ) } + 1 ]^{-1}
 $
is the Fermi function at inverse temperature $\beta$ and chemical potential $\mu$, and 
we have taken the limit $V \rightarrow \infty$ to replace the sum over $\bd{q}$ by an integral. 
The first-order Hartree and Fock diagrams taken into account in Eq.~(\ref{eq:sigma1}) are shown in Fig.~\ref{fig:diagrams} (a).
Since the limit ${\bd{q}} \rightarrow 0 $ of $ U_{\bd{q}}$ is ambiguous, 
we follow Fregoso {\it{et al.}}~\cite{Fregoso2009} and define $U_0$ in terms of the angular 
average of $U_{\bd{q}}$. This amounts to formally setting $U_0  \rightarrow 0$ so that 
all Hartree bubbles vanish. We may improve the first-order  approximation by replacing 
$\epsilon_{\bd{k} + \bd{q}} \rightarrow \epsilon_{\bd{k} + \bd{q}} + 
\Sigma ( \bd{k} + \bd{q} )$ 
on the right-hand side of Eq.~(\ref{eq:sigma1}), so that we obtain
 \begin{equation}
 \Sigma^{\rm HF} ( \bd{k} ) = \int \frac{ d^3 q}{(2 \pi)^3}
 ( U_0 - U_{ \bd{q}}  ) f \left( \epsilon_{\bd{k} + \bd{q}} + \Sigma^{\rm HF} ( \bd{k} + \bd{q} ) \right),
 \label{eq:sigma1b}
 \end{equation}
known as the self-consistent Hartree-Fock approximation. Diagrammatically, this equation amounts 
to an infinite resummation of perturbation theory, where the bare propagators in the first-order diagrams shown in Fig.~\ref{fig:diagrams}~(a) are replaced by self-consistent propagators
 \begin{equation}
 G^{\rm HF} ( \bd{k} , \omega  ) = \frac{1}{  \omega - \epsilon_{\bd{k}} + \mu - 
 \Sigma^{\rm HF} ( \bd{k} )}.
 \end{equation}

The angular dependence of  the first-order self-energy $\Sigma^{(1)} ( \bd{k} )$
can be extracted analytically by means of a suitable rotation of the integration
variables. Since we shall use the same procedure for the evaluation of the
second-order self-energy, let us explain this in some detail.
The dependence of the integral in Eq.~(\ref{eq:sigma1})
on the angular part $\hat{\bd{k}}$ of $\bd{k}$ enters in the form
 \begin{equation}
  I ( \hat{\bd{k}} , \hat{\bd{d}} ) = \int \frac{ d^3 q}{(2 \pi)^3}
 A ( \hat{\bd{d}} \cdot \hat{\bd{q}}    )
 B ( \hat{\bd{k}} \cdot {\bd{q}} ),
 \label{eq:IAB}
 \end{equation}
where the functions  $A$ and $B$ denote the different factors of the integrand resulting 
from the substitution of Eq.~(\ref{eq:fourier_transform_3d}) into Eq.~(\ref{eq:sigma1}). Let us now define 
rotated integration variables
 $\bd{p} = e^{ \bd{\alpha} \times } \bd{q},  $
where we have represented the rotation of a vector $\bd{q}$ around an axis
$\hat{\bd{\alpha}} = \bd{\alpha} / | \bd{\alpha} |$  with angle
$\alpha = | \bd{\alpha} |$ in terms of an exponentiated cross product,
 \begin{equation}
  e^{ \bd{\alpha} \times } \bd{q} = 
 \hat{\bd{\alpha}} ( \hat{\bd{\alpha}} \cdot \bd{q} )
 +  \hat{\bd{\alpha}} \times \bd{q}\sin \alpha
 -  \hat{\bd{\alpha}} \times ( \hat{\bd{\alpha}} \times \bd{q} ) \cos \alpha.
  \end{equation}
Choosing $\bd{\alpha}$ such that the corresponding rotation maps the
direction  $\hat{\bd{k}} $
into the direction $\hat{\bd{d}}$ of the dipoles (see Fig.~\ref{fig:dipole3}), we have
 \begin{equation}
 \hat{\bd{d}} =  e^{ \bd{\alpha} \times } \hat{\bd{k}}, \; \; \; \hat{\bd{\alpha}} =
 \frac{ \hat{\bd{k}} \times \hat{\bd{d}} }{ | \hat{\bd{k}} \times \hat{\bd{d}} | },
 \end{equation}
where
$ \cos \alpha = k_{\parallel} = \hat{\bd{k}} \cdot
 \hat{\bd{d}}.$
We may always choose the angle $\alpha$ such that $0 \leq \alpha \leq \pi$ and therefore
$ \sin \alpha = k_{\bot} = | \hat{\bd{k}} \times
 \hat{\bd{d}} |.$
Next, we use the invariance of the scalar products under rotations 
and obtain
\begin{equation}
  I ( \hat{\bd{k}} , \hat{\bd{d}} ) = \int \frac{ d^3 p}{(2 \pi)^3}
 A \left( (e^{\bd{\alpha} \times} \hat{\bd{d}} ) \cdot \hat{\bd{p}} \right)  
 B ( \hat{\bd{d}} \cdot {\bd{p}}  ).
 \end{equation}
With these definitions
$ e^{\bd{\alpha} \times} \hat{\bd{d}} =  \hat{\bd{d}} ( \hat{\bd{d}} \cdot
 \hat{\bd{k}} ) - {\bd{k}}_{\bot},$
where
${\bd{k}}_{\bot} = \hat{\bd{k}} - \hat{\bd{d}} ( \hat{\bd{d}} \cdot \hat{\bd{k}} )$
is the component of the unit vector $\hat{\bd{k}}$ perpendicular
to $\hat{\bd{d}}$ as shown in Fig.~\ref{fig:dipole3}.
We obtain
 \begin{equation}
  I ( \hat{\bd{k}}, \hat{\bd{d}} ) = \int \frac{ d^3 p}{(2 \pi)^3}
 A \left(    ( \hat{\bd{d}} \cdot \hat{\bd{p}} )    ( \hat{\bd{d}} \cdot \hat{\bd{k}} ) 
 -  {\bd{p}}_{\bot} \cdot {\bd{k}}_{\bot}  \right)  
 B ( \hat{\bd{d}} \cdot {\bd{p}}  ),
 \label{eq:I3}
 \end{equation}
where again
${\bd{p}}_{\bot} = \hat{\bd{p}} - \hat{\bd{d}} ( \hat{\bd{d}} \cdot \hat{\bd{p}} )$.
Using the coordinate system and the spherical coordinates defined
in Fig.~\ref{fig:dipole3} we arrive at the following expression for the integral (\ref{eq:IAB}),
  \begin{eqnarray}
  I ( \hat{\bd{k}}, \hat{\bd{d}} ) 
 & = & \frac{1}{(2 \pi)^3} \int_0^{\infty} dp  p^2  \int_0^{\pi} d \theta \sin \theta 
 \int_0^{2 \pi} d \varphi  
 \nonumber
 \\*
 & \times & A ( k_{\parallel} \cos \theta - k_{\bot} 
 \sin \theta \cos \varphi ) B ( p \cos \theta ).
 \hspace{7mm}
 \end{eqnarray}
Applying this to Eq.~(\ref{eq:sigma1}) we find
 \begin{eqnarray}
 \Sigma^{(1)} ( \bd{k} ) & = &  - \frac{8 \pi d^2}{3 ( 2 \pi )^2 }
 \int_0^{\infty} dp  p^2  \int_0^{\pi} d \theta \sin \theta 
 \nonumber
 \\*
 & \times &
 \int_0^{2 \pi} \frac{d \varphi}{2 \pi}  
 P_2 ( k_{\parallel} \cos \theta - k_{\bot} 
 \sin \theta \cos \varphi )
 \nonumber
 \\*
 & \times & f \left( \epsilon_k + \epsilon_p + \frac{k p}{m} \cos \theta \right).
 \end{eqnarray}
The $\varphi$ integration can now be performed,
 \begin{equation}
 \int_0^{2 \pi} \frac{d \varphi}{2 \pi}  
 P_2 ( k_{\parallel} \cos \theta - k_{\bot} 
 \sin \theta \cos \varphi ) = P_2 ( \cos \alpha ) P_2 ( \cos \theta ),
 \end{equation}
where we have used $\cos \alpha = k_{\parallel} = \hat{\bd{k}} \cdot \hat{\bd{d}}$.
It is convenient to  introduce the dimensionless coupling constant
 \begin{equation}
 u = \nu \frac{8 \pi d^2}{3} = 4 \pi \frac{n d^2}{E_{F0}} = \frac{4 d^2 m k_{F0}}{3 \pi},
 \end{equation}
where $E_{F0} = k_{F0}^2 / (2m)$ is the Fermi energy of the noninteracting 
system, $\nu = m k_{F0} / (2 \pi^2)$ is its density of states at the Fermi energy,
and $n = k_{F0}^3 / (6 \pi^2)$ is the particle density in three dimensions.
In the limit of vanishing temperature we obtain
 \begin{eqnarray}
 \frac{ \Sigma^{(1)} ( \bd{k} )}{ \mu } & = & -  \rFactor u P_2 ( \cos \alpha )
 \int_0^{\infty} \frac{ dp p^2}{(\rFactor k_{F0})^3} 
 \int_0^{\pi} d \theta \sin \theta 
 \nonumber
 \\*
 & \times & P_2 ( \cos \theta ) \Theta \left( \mu - \epsilon_k - \epsilon_p - \frac{k p}{m} \cos \theta \right),
 \label{eq:Sigma1a}
 \end{eqnarray}
where $\Theta (x)$ represents the Heaviside step function. 
The factor $\rFactor = \rFactor(u)$ is defined 
in terms of the ratio between the renormalized chemical potential and the bare Fermi energy,
\begin{equation}
\label{eq:def_r_factor}
 \rFactor^2 = \frac{\mu}{ E_{F0}} =  \frac{\mu}{  k_{F0}^2 / (2m) }.
\end{equation}
As usual, we work at fixed particle density, so that the value of $\mu$ should be adjusted
to keep the density constant when the interaction is switched on.
The explicit calculation of the renormalized chemical potential
will be discussed in Sec.~\ref{sec:results_fermi_surface} where
we shall show that $\gamma = 1 -0.10 u^2 + {\cal{O}} ( u^3 )$
[see Eq.~(\ref{eq:r_factor_result})], so that to first order in the interaction we may set $\gamma \approx 1$.
For convenience  we introduce the dimensionless variables
$ \tilde{k} =  k /(\rFactor k_{F0} )$ and $\tilde{p} = p / ( \rFactor k_{F0} )$;  the integrand 
in Eq.~(\ref{eq:Sigma1a}) is then independent of $\mu$. Performing the remaining integrations 
one  finally  obtains~\cite{Cha10,Fregoso2010}
\begin{equation}
\label{eq:firstOrderResult}
\frac{\Sigma^{(1)} (\bd{{k}})}{\mu} = - \frac{ \rFactor u }{3}  H^{(1)} (\tilde{k}) P_2 (\hat{\bd{k}} \cdot
 \hat{\bd{d}} ),
\end{equation}
where
\begin{eqnarray}
\label{eq:analytFirstOrder}
H^{(1)} (\tilde{k}) & = & \frac{1}{8 \tilde{k}^3} \biggl[ - 3 \tilde{k} + 8 \tilde{k}^3 + 3 \tilde{k}^5 - \frac{3}{2} \bigl(\tilde{k}^2 - 1\bigr)^3 \ln \Bigl|\frac{\tilde{k} + 1}{\tilde{k} - 1}\Bigr|\biggr]
\nonumber
\\*
\end{eqnarray}
%
is a positive, continuous function with $H^{(1)} (1) = 1$. The first-order result (\ref{eq:firstOrderResult})
 implies that to lowest 
order in the interaction the Fermi surface is distorted in the direction of the external field 
which reflects the anisotropy of the interaction. Moreover, the renormalized
Fermi velocity acquires an angular dependence. 
We postpone a more detailed discussion 
to  Sec.~\ref{sec:results} where we shall also discuss the 
results obtained in  second-order perturbation theory.

\section{Second-order self-energy}
 \label{sec:second}

To second order in the interaction, there are totally six diagrams contributing to the
self-energy which we show in  Fig.~\ref{fig:diagrams}~(b) and (c).
The four diagrams in the group (b)
are  implicitly taken into account via the Hartree-Fock self-consistency 
condition in Eq.~(\ref{eq:sigma1b}). The contribution of these diagrams
can therefore be calculated analytically;  together with  the first-order contribution 
(\ref{eq:analytFirstOrder}) we obtain at this level of approximation
%
\begin{eqnarray}
\label{eq:sigmaHF}
\frac{\Sigma^{\rm HF } (\bd{{k}})}{\mu} = & - & \frac{ \rFactor u}{3}  H^{(1)} (\tilde{k}) 
P_2 (  \hat{\bd{k}} \cdot \hat{\bd{d}}    ) + (\rFactor u)^2 \Bigl[ - \frac{1}{60} H^{(2)}_0 (\tilde{k})
\nonumber
\\*
 & + & \frac{1}{42} H^{(2)}_2 (\tilde{k}) P_2 (  \hat{\bd{k}} \cdot \hat{\bd{d}}  ) - \frac{1}{140} 
H^{(2)}_4 (\tilde{k}) P_4 (  \hat{\bd{k}} \cdot \hat{\bd{d}} ) \Bigr]
\nonumber
\\*
& + & {\cal{O}} ( u^3 )
, \hspace{7mm}
\end{eqnarray}
where $P_4 ( x ) = \frac{1}{8} [ 35 x^4 -30 x^2 + 3 ]$ is the
fourth Legendre polynomial and 
the functions $H^{(2)}_n ( \tilde{k}) =    H^{(2)}_n (  k /( \gamma k_{F0} ))$ are given  by
\begin{subequations}
\begin{eqnarray}
H^{(2)}_0 (\tilde{k}) & = & \frac{1}{4 \tilde{k}} \Bigl[ 10 \tilde{k} - 6 \tilde{k}^3 + 3 \bigl(\tilde{k}^2 - 1\bigr)^2 \ln \Bigl|\frac{\tilde{k} + 1}{\tilde{k} - 1}\Bigr|\Bigr],
\\
H^{(2)}_2 (\tilde{k}) & = & \frac{1}{8 \tilde{k}^3} \Bigl[ 6 \tilde{k} - 4 \tilde{k}^3 + 6 \tilde{k}^5 - 3 \bigl( \tilde{k}^6 - \tilde{k}^4 - \tilde{k}^2 + 1 \bigr)
\nonumber
\\*
& \times & \ln \Bigl|\frac{\tilde{k} + 1}{\tilde{k} - 1}\Bigr|\Bigr],
\\
H^{(2)}_4 (\tilde{k}) & = & \frac{1}{32 \tilde{k}^5} \Bigl[ - 210 \tilde{k} + 290 \tilde{k}^3 - 30 \tilde{k}^5 - 18 \tilde{k}^7
\nonumber
\\*
& + & 3 \bigl(\tilde{k}^2 - 1\bigr)^2 \bigl( 3 \tilde{k}^4 + 10 \tilde{k}^2 + 35 \bigr) \ln \Bigl|\frac{\tilde{k} + 1}{\tilde{k} - 1}\Bigr|\Bigr].
\nonumber
\\*
\end{eqnarray}
\end{subequations}
%
Note that these functions are positive and  continuous  with $H^{(2)}_n (1) = 1$.

To complete the second-order calculation,
we should add the contribution from the two diagrams shown in Fig.~\ref{fig:diagrams}~(c) 
which are not taken into account in self-consistent mean-field theory; 
the total self-energy to second order in the interaction is then given by 
 \begin{equation}
 \Sigma ( \bd{k}, i \omega ) = \Sigma^{\rm{HF}} ( \bd{k} ) + \Sigma^{(2)} ( \bd{k} , i \omega ) +
 {\cal{O}} ( u^3 ),
 \end{equation} 
where $\Sigma^{\rm{HF}} ( \bd{k} )$ was defined in Eq.~(\ref{eq:sigmaHF}) and $\Sigma^{(2)} ( \bd{k} , i \omega )$, representing the
contribution of the two diagrams in Fig.~\ref{fig:diagrams}~(c), is given by
\begin{eqnarray}
\label{eq:sigma2}
 \Sigma^{(2)} ( \bd{k}, i \omega  ) & = & - \frac{1}{(\beta V)^2}  \sum_{Q ,Q^{\prime}} U_{\bd{q}} [    U_{\bd{q}}  -  U_{\bd{q}^{\prime} }  ] G_0 ( K + Q )
 \nonumber
 \\*
 & \times & G_0 ( K + Q^{\prime} ) G_{0} ( K + Q + Q^{\prime} ),
\end{eqnarray}
where in the right-hand side
$K = ( \bd{k} , i \omega )$ is a collective label for momentum $\bd{k}$ and fermionic Matsubara frequency
$i \omega$,  while $Q = ( \bd{q} , i \bar{\omega} )$ and  
$Q^{\prime} = ( \bd{q}^{\prime} , i \bar{\omega}^{\prime} )$ depend on bosonic Matsubara frequencies $i \bar{\omega}$ and $i \bar{\omega}^{\prime}$.
Moreover,
$G_0  ( K) = [ i \omega - \epsilon_{\bd{k}} + \mu ]^{-1}$ 
is the  noninteracting Matsubara Green's function.  
The frequency sums in Eq.~(\ref{eq:sigma2}) can be easily carried out.
To obtain the retarded self-energy, we then
perform the analytic continuation $i\omega \rightarrow \omega + i0^+$ 
to real frequencies and obtain in the limit of 
vanishing temperature and infinite volume
 \begin{widetext}
 \begin{eqnarray}
 \Sigma^{(2)} ( \bd{k} , \omega  ) & = &   \int \frac{ d^3 q}{(2 \pi)^3}
 \int \frac{ d^3 q^{\prime}}{(2 \pi )^3}
  \frac{1}{2} [    U_{\bd{q}}  -  U_{\bd{q}^{\prime} }  ]^{2}
 \frac{  \Theta ( \xi_{ \bd{k} + \bd{q}} )  \Theta ( \xi_{ \bd{k} + \bd{q}^{\prime}} )
  \Theta ( - \xi_{ \bd{k} + \bd{q} + \bd{q}^{\prime}} )  +
\Theta ( - \xi_{ \bd{k} + \bd{q}} )  \Theta ( - \xi_{ \bd{k} + \bd{q}^{\prime}} )
  \Theta (  \xi_{ \bd{k} + \bd{q} + \bd{q}^{\prime}} )
  }{\omega + i0^+ -  ( \xi_{ \bd{k} + \bd{q}} +  \xi_{ \bd{k} + \bd{q}^{\prime}} 
  - \xi_{ \bd{k} + \bd{q} + \bd{q}^{\prime}}         )     },
 \hspace{7mm}
 \label{eq:sigma2b}
 \end{eqnarray}
%
where $\xi_{\bd{k}} = \epsilon_{\bd{k}} - \mu$ and 
we have rewritten the interaction so that the integrand is manifestly symmetric 
under  ${\bd{q}} \leftrightarrow {\bd{q}^{\prime}}$. 
The dependence of $\Sigma^{(2)} ( \bd{k} , \omega  ) $ 
on the angular part of ${\bd{k}}$ can be extracted analytically by rotating the integration 
variables $\bd{q}$ and $\bd{q}^{\prime}$ as described in Sec.~\ref{sec:firstorder};
i.e., we introduce $\bd{p} = e^{\bd{\alpha} \times } \bd{q}$ and
$\bd{p}^{\prime} = e^{\bd{\alpha} \times } \bd{q}^{\prime}$,
where the rotation matrix $e^{ \bd{\alpha} \times}$ rotates
$\hat{\bd{k}}$ into $\hat{\bd{d}}$.
After these transformations we obtain
 \begin{eqnarray}
 \frac{\Sigma^{(2)} ( \bd{k} , \omega )}{\mu} & = & (\rFactor u)^2 \Big[ \Sigma^{(2)}_0 ( k , \omega )
 + P_2 ( \hat{\bd{k}} \cdot \hat{\bd{d}} )  \Sigma^{(2)}_2 ( k , \omega )
+ P_4 (  \hat{\bd{k}} \cdot \hat{\bd{d}}    )  \Sigma^{(2)}_4 ( k , \omega )\Big],
 \end{eqnarray}
with
 \begin{eqnarray}
  \Sigma_{n}^{(2)} ( k , \omega )& = &
 \int_0^{\infty} \frac{ dp p^2}{(\rFactor k_{F0})^3} 
\int_0^{\infty} \frac{ dp^{\prime} p^{\prime 2}}{(\rFactor k_{F0})^3}
 \int_0^{\pi} d \theta \sin \theta \int_0^{\pi} d \theta^{\prime} \sin \theta^{\prime}
 \int_0^{2 \pi} \frac{ d \phi}{2 \pi}  A_n ( \theta , \theta^{\prime} , \phi )
 \nonumber
 \\*
 & \times & \mu
 \frac{  \Theta ( \xi_{ k \hat{\bd{d}} + \bd{p}} )  \Theta ( \xi_{ k \hat{\bd{d}} + \bd{p}^{\prime}} )
  \Theta ( - \xi_{ k \hat{\bd{d}} + \bd{p} + \bd{p}^{\prime}} )  +
\Theta ( - \xi_{ k \hat{\bd{d}} + \bd{p}} )  \Theta ( - \xi_{ k \hat{\bd{d}} + \bd{p}^{\prime}} )
  \Theta (  \xi_{ k \hat{\bd{d}} + \bd{p} + \bd{p}^{\prime}} )
  }{\omega + i0^+ -  ( \xi_{ k \hat{\bd{d}} + \bd{p}} +  \xi_{ k \hat{\bd{d}} + \bd{p}^{\prime}} 
  - \xi_{ k \hat{\bd{d}} + \bd{p} + \bd{p}^{\prime}} ) },
 \end{eqnarray}
where $\bd{p} \cdot \bd{p}^{\prime} = p p^{\prime} [ 
\cos \theta \cos \theta^{\prime} + \sin \theta \sin \theta^{\prime} \cos \phi]$.
The coefficients $A_n ( \theta , \theta^{\prime} , \phi )$ are defined 
via the expansion
 \begin{eqnarray}
 A ( \alpha ; \theta , \theta^{\prime} , \phi ) & \equiv & \int_0^{2 \pi} \frac{ d \varphi}{2 \pi} \frac{1}{2}
 \Bigl[ P_2 ( \cos \alpha \cos \theta 
 - \sin \alpha \sin \theta \cos \varphi )
-  P_2 ( \cos \alpha \cos \theta^{\prime} 
 - \sin \alpha \sin \theta^{\prime} \cos (\varphi + \phi) ) 
 \Bigr]^{2}
 \nonumber
 \\*
 & = & A_0 ( \theta , \theta^{\prime} , \phi ) 
+ P_2 ( \cos \alpha )  A_2 ( \theta , \theta^{\prime} , \phi )
+ P_4 ( \cos \alpha )  A_4 ( \theta , \theta^{\prime} , \phi )
 \end{eqnarray}
and are explicitly given by
 \begin{subequations}
 \begin{eqnarray}
  A_0 ( \theta , \theta^{\prime} , \phi ) & = & -\frac{3}{80} \Bigl[-5 + 2 \cos (\theta + \theta^{\prime}) \cos (\theta - \theta^{\prime}) + 3 \cos (2\theta) \cos (2\theta^{\prime}) + 4 \sin (2\theta) \sin (2\theta^{\prime}) \cos \phi
 \nonumber
 \\*
 & + & 4 \sin^{2} \theta \sin^{2} \theta^{\prime} \cos (2\phi)\Bigl],
 \\
 A_2 ( \theta , \theta^{\prime} , \phi ) & = & \frac{3}{56} \Bigl[1 + 2 \cos (\theta + \theta^{\prime}) \cos (\theta - \theta^{\prime}) - 3 \cos (2\theta) (2\theta^{\prime}) - 2 \sin (2\theta) \sin (2\theta^{\prime}) \cos \phi
 \nonumber
 \\*
 & + & 4 \sin^{2} \theta \sin^{2} \theta^{\prime} \cos (2\phi)\Bigl],
 \\
 A_4 ( \theta , \theta^{\prime} , \phi ) & = & -\frac{9}{1120} \Bigl[-5 + 4 \cos (\theta + \theta^{\prime}) \cos (\theta - \theta^{\prime}) - 35 \cos (2\theta + 2\theta^{\prime}) \cos (2\theta - 2\theta^{\prime}) + 36 \cos (2\theta) \cos (2\theta^{\prime})
 \nonumber
 \\*
 & - & 32 \sin (2\theta) \sin (2\theta^{\prime}) \cos \phi + 8 \sin^{2} \theta \sin^{2} \theta^{\prime} \cos (2\phi)\Bigl].
 \end{eqnarray}
 \end{subequations}
The factor $\rFactor$ is defined in Eq.~(\ref{eq:def_r_factor}).
As in Sec.~\ref{sec:firstorder}
we now introduce dimensionless momenta $  \tilde{k}  =k/ (\rFactor k_{F0}) $, 
$\tilde{p} = p/ (\rFactor k_{F0}) $, and $\tilde{p}^{\prime} = p^{\prime} / (\rFactor k_{F0})$, 
as well as the dimensionless frequency $\tilde{\omega}  = \omega/ \mu $.
The $\mu$ dependence of the integrand can then be scaled out and the  $\tilde{p}^{\prime}$ integration can be performed analytically,
with the result
%
\begin{eqnarray}
  \Sigma_{n}^{(2)} ( {\tilde{k}} ,  {\tilde{\omega}} )& = &
  \int_0^{\infty} d\tilde{p} \tilde{p}^2
 \int_0^{\pi} d \theta \sin \theta \int_0^{\pi} d \theta^{\prime} \sin \theta^{\prime}
 \int_0^{2 \pi} \frac{ d \phi}{2 \pi}  A_n ( \theta , \theta^{\prime} , \phi )
 \nonumber
 \\*
& \times & \left[ \Theta \bigl( \tilde{k}^2 + \tilde{p}^2 + 2 \tilde{k} \tilde{p} \cos \theta -1 \bigl) 
Q_{12} ( \tilde{k} , \tilde{\omega};  \tilde{p}, \theta, \theta^{\prime}, \phi ) 
+ \Theta \bigl(  1 - \tilde{k}^2 - \tilde{p}^2 -  2 \tilde{k} \tilde{p} \cos \theta     \bigl) Q_{21} ( \tilde{k} , \tilde{\omega};  \tilde{p}, \theta, \theta^{\prime}, \phi ) \right].
 \hspace{7mm}
 \label{eq:selffour}
 \end{eqnarray}
Here the functions
$Q_{ij} ( \tilde{k} , \tilde{\omega};  \tilde{p}, \theta, \theta^{\prime}, \phi ) $ (with $ij=12$ 
or $ij=21$) are given by
\begin{eqnarray}
 & &  Q_{{ij}}   ( \tilde{k} , \tilde{\omega};  \tilde{p}, \theta, \theta^{\prime}, \phi )  
= \Theta (- r_{i}) \Theta (r_{j}) \Theta (p^{+}_{j} - 
 m^a_{i}) F(m^a_{i}, p^+_{j})
 + \Theta (r_{i}) \Theta (r_{j}) \Big[ \Theta (p^{+}_{j} 
- m^b_{i}) F(m^b_{i}, p^+_{j})
 + \Theta (m^c_{i} - m^a_{i}) F(m^a_{i}, 
 m^c_{i}) \Big],
 \nonumber
 \\*
\end{eqnarray}
\end{widetext}
where we have introduced the abbreviations
\begin{subequations}
\begin{eqnarray}
 p^{\pm}_1 &=& -\tilde{k} \cos \theta^{\prime} \pm \sqrt{r_1},
 \\
 p^{\pm}_2 &=& -\tilde{k} \cos \theta^{\prime} - \tilde{p} w \pm \sqrt{r_2},
 \\
 w &=& \cos \theta \cos \theta^{\prime}
 + \sin \theta \sin \theta^{\prime} \cos \phi,
 \\
 r_{1} &=& 1 - \tilde{k}^2 \sin^2 \theta^{\prime},
 \\
 r_{2} &=& (\tilde{k} \cos \theta^{\prime} + \tilde{p} w )^2 - \tilde{p}^2 -
 2 \tilde{k} \tilde{p} \cos \theta - \tilde{k}^2 + 1,
 \hspace{7mm}
 \\
 m^a_{1} &=& \max(0,p^{-}_{2}), \; \; \; \; \; \; \, \; \; m^a_{2} = \max(0,p^{-}_{1}),
 \\
 m^b_{1} &=& \max(0,p^{-}_{2},p^{+}_{1}), \; \; \; m^b_{2} 
= \max(0,p^{-}_{1},p^{+}_{2}),
 \\
 m^c_{1} &=& \min(p^{-}_{1},p^{+}_{2}), \; \; \; \; \; \; \, m^c_{2} = \min(p^{-}_{2},p^{+}_{1}),
\end{eqnarray}
\end{subequations}
and the function
\begin{eqnarray}
 F(x,y) &=& - \frac{x-y}{ 4 \tilde{p}^2 w^2 } 
\left[\tilde{p} w (x+y) - \tilde{\omega} + \tilde{k}^2 -1 \right]
 \nonumber
 \\*
 &  & \hspace{-20mm} + \frac{ ( \tilde{\omega} - \tilde{k}^2 + 1 )^2 }{8 \tilde{p}^3 w^3}
 \ln \left[ \frac{\tilde{\omega} + i0^+ - \tilde{k}^2 + 1 + 2 \tilde{p} w y }{ 
 \tilde{\omega} + i0^+ - \tilde{k}^2 + 1 + 2 \tilde{p} w x} \right].
\end{eqnarray}
By splitting the
complex function $F (x,y)$ into its real and imaginary part,
we obtain the  real and the imaginary part of the second-order self-energy.
We have performed the  remaining four-dimensional integration
in Eq.~(\ref{eq:selffour})  numerically using the VEGAS Monte Carlo algorithm 
from the GNU Scientific Library~\cite{GSL}.

\section{Renormalized Fermi surface and quasiparticle properties}
\label{sec:results}

\subsection{General definitions}

Given the momentum and frequency dependent 
retarded self-energy $\Sigma ( \bd{k} , \omega )$, the wavevectors 
on the  renormalized Fermi surface can be obtained from the solution of
 \begin{equation}
 \epsilon_{\bd{k}_F} + \Sigma ( \bd{k}_F , i0^+ ) = \mu.
 \label{eq:FSdef}
 \end{equation}
Moreover the effective mass and the quasiparticle residue can be defined 
in terms of the low-energy expansion of the self-energy around the
renormalized Fermi surface,
 \begin{eqnarray}
 \Sigma ( \bd{k}_F + \bd{q} , \omega ) & \approx & \Sigma ( \bd{k}_F , i0^+ ) +
 \left. \nabla_{\bd{k}}  \Sigma ( \bd{k} , i0^+ ) \right|_{ \bd{k} = \bd{k}_F } 
 \cdot \bd{q}
 \nonumber
 \\*
 & + &
 \left.
 \frac{ \partial  \Sigma ( \bd{k}_F , \omega ) }{\partial \omega }
 \right|_{ \omega = i0^+ } \omega .
 \end{eqnarray}
In this approximation the retarded Green's function has the quasiparticle form
 \begin{equation}
  G ( \bd{k}_F + \bd{q} , \omega)
 \approx \frac{ Z_{\bd{k}_F }}{ \omega + i0^+ - {\bd{v}}_{\bd{k}_F} \cdot \bd{q} },
 \end{equation}
with the quasiparticle residue
 \begin{equation}
 \label{eq:qp_res_def}
 Z_{\bd{k}_F} = \frac{1}{1 - 
 \left. \frac{ \partial \Sigma ( \bd{k}_F , \omega   )  }{ \partial \omega }
 \right|_{ \omega = i0^+} }
 \end{equation}
and the renormalized Fermi velocity
 \begin{equation}
 \label{eq:vf_def}
  {\bd{v}}_{\bd{k}_F} = Z_{\bd{k}_F} 
 \left[ \frac{ \bd{k}_F }{m} +    
 \left. \nabla_{\bd{k}}  \Sigma ( \bd{k} , i0^+ ) \right|_{ \bd{k} = \bd{k}_F } 
 \right].
 \end{equation}
Note that by construction  $ {\bd{v}}_{\bd{k}_F}$ is perpendicular to the
renormalized Fermi surface at point $\bd{k}_F$. Therefore the information about the 
direction of the Fermi velocity is redundant if we know the shape of the
renormalized Fermi surface and
we can restrict ourselves to the calculation of $  |{\bd{v}}_{\bd{k}_F}|$.
The effective mass can be defined by setting
 \begin{equation}
 | \bd{v}_{\bd{k}_F} |  = \frac{ | \bd{k}_F | }{ m^{\ast}},
 \end{equation}
but this obviously does not contain any new information beyond  $| \bd{k}_F |$ and
$| \bd{v}_F |$.
Because in second-order perturbation theory the self-energy has also an imaginary part,
we obtain a broadened spectral function
\begin{equation}
\label{eq:spectral_def}
\rho (\bd{k}, \omega) = - \frac{1}{\pi} \frac{\text{Im} \Sigma (\bd{k}, \omega)}{\left[ \omega - \epsilon_{\bd{k}} + \mu - \text{Re} \Sigma (\bd{k}, \omega) \right]^2 + \left[ \text{Im} \Sigma (\bd{k}, \omega) \right]^2}.
\end{equation}

\subsection{Renormalized Fermi surface}
\label{sec:results_fermi_surface}

To begin with, let us calculate the renormalized Fermi surface,
which we parametrize by $\bd{k}_F = k_F ( \alpha )  \hat{\bd{k}}_F $,
where $\alpha$ is the angle between $\bd{k}_F$ and the direction
$\hat{\bd{d}}$ of the dipoles; i.e., $\cos \alpha = \hat{\bd{k}}_F \cdot \hat{\bd{d}}$.
Substituting the definition
$\mu = \gamma^2 E_{F0}$ introduced in Eq.~(\ref{eq:def_r_factor})
into the defining equation (\ref{eq:FSdef}) of the renormalized Fermi surface 
we obtain
\begin{equation}
 \frac{k_F (\alpha)}{\rFactor k_{F0}} \equiv \tilde{k}_F (\alpha) = 
\sqrt{1 - \frac{\Sigma ( \tilde{k}_F (\alpha) \hat{\bd{k}}_F   ,i0^+ )}{\mu}}. 
 \label{eq:k_f_alpha}
\end{equation}
Given our perturbative result
for the self-energy we can now iterate Eq.~(\ref{eq:k_f_alpha}) to obtain
an expansion of $\tilde{k}_F ( \alpha )$ in powers of $(\rFactor u)$. 
Since we  keep the particle density $n$ fixed, 
Luttinger's theorem~\cite{Luttinger1960a} tells us that the volume of the Fermi surface must not change due to the interaction, so that we can fix the factor $\rFactor$ from
the condition
\begin{eqnarray}
\label{eq:luttinger}
\left( \rFactor k_{F0} \right)^3 2 \pi   \int_0^{\pi} d \alpha \sin \alpha   \int_0^{ 
 \tilde{k}_F ( \alpha) } d \tilde{k} \tilde{k}^2   = \frac{4 \pi}{3} k_{F0}^3,
\end{eqnarray}
where we have introduced the 
rescaled integration variable $\tilde{k} = k /(\rFactor k_{F0} )$. 
Substituting the perturbative expression for $\tilde{k}_F (\alpha)$ from 
Eq.~(\ref{eq:k_f_alpha}) into Eq.~({\ref{eq:luttinger}) and expanding the integral to
second order in $(\rFactor u)$, we determine $\rFactor$ to second order in the
interaction as
\begin{equation}
\label{eq:r_factor_result}
\rFactor = 1 - 0.10 u^2 + {\cal{O}} ( u^3).
\end{equation}
The reason why there is no term linear in  $u$ is that the first-order self-energy 
is proportional to $P_2 (\cos \alpha)$ [see Eq.~(\ref{eq:firstOrderResult})], so that
the corresponding first-order contribution to the integral over the 
Fermi volume vanishes. 
From Eq.~(\ref{eq:r_factor_result}) we may then determine the renormalized Fermi surface to second order,
\begin{eqnarray}
\frac{k_F (\alpha)}{k_{F0}} &=& 1 + \frac{u}{6} P_2 (\cos \alpha) - u^2 \Bigl[ \frac{1}{180} - 0.031 P_2 (\cos \alpha)
\nonumber
\\*
&-& 0.016 P_4 (\cos \alpha) \Bigl] + {\cal{O}} ( u^3 ).
\label{eq:k_f_alpha_num}
\end{eqnarray}
The corresponding Fermi surface for 
$u = 1.5$
is shown in Fig.~\ref{fig:fermi_surface}. 
\begin{figure}
\includegraphics[width=\linewidth]{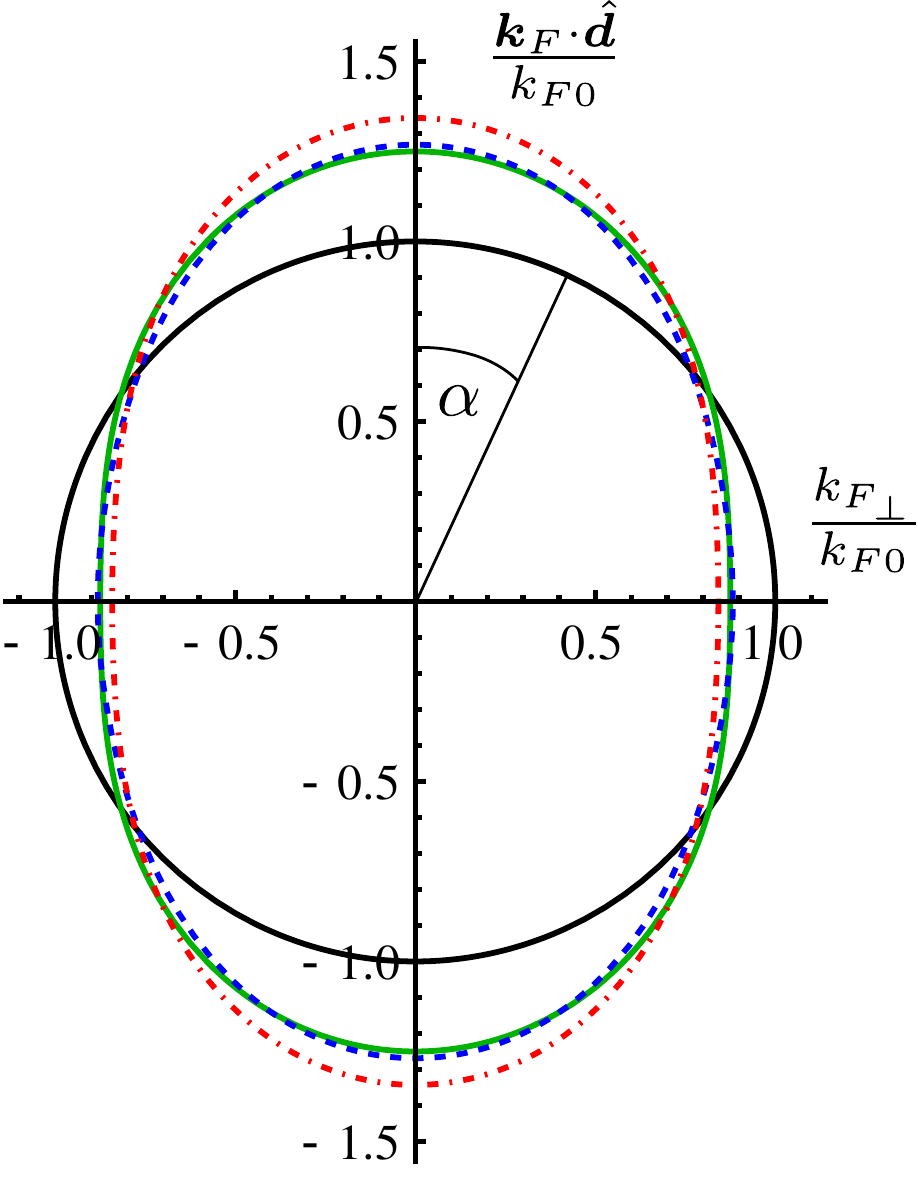}
\caption{
(Color online)
Fermi surface of dipolar fermions for $u = 1.5$ 
to first order (solid green), to second 
order in the self-consistent Hartree-Fock approximation (dashed blue),
 and to full second order (dot-dashed red) in the interaction. 
The spherical Fermi surface of the noninteracting system is given as a reference 
(solid black line). 
Note that the deformed Fermi surface still has the azimuthal symmetry around the $z$ axis.
}
\label{fig:fermi_surface}
\end{figure}
Such a large value of the interaction  is close to the stability limit of the
Fermi liquid state  (see Sec.~\ref{subsec:stab_limit});  in the weak-coupling limit
 $u \ll 1$ (where our perturbative calculation 
can be trusted) the second-order correction is
barely visible. From Fig.~\ref{fig:fermi_surface} we see
that the second-order correction enhances the 
tendency found in  the first-order calculation to distort the Fermi surface
along the direction of the dipoles.
We also see that the true many-body corrections to the self-energy
shown in Fig.~\ref{fig:diagrams} (c) have a much stronger effect
than the second-order diagrams taken into account 
via the  self-consistent Hartree-Fock approximation.
To make contact with the recent experiment by 
Aikawa {\it{et al.}} [\onlinecite{aikawa2014observation}], we show in
Fig.~\ref{fig:aspect_ratio} how
the aspect ratio of the Fermi surface, defined by $k_F (0) / k_F (\frac{\pi}{2})$, changes
in the experimentally relevant range of interactions.
Obviously, the second-order correction leads to a slightly 
larger deviation from the spherical shape of the Fermi surface, 
but for the experimentally relevant range of interactions
the second-order correction is more than an order of
magnitude smaller than the first-order result.
Hence, for the range of interactions relevant to
the experiment by Aikawa {\it{et al.}} [\onlinecite{aikawa2014observation}] the deformation of the Fermi surface can be accurately
calculated in first order perturbation theory. However, in two dimensions Fermi surface deformations can go first order in a non-analytic way \cite{Yamaji2006,Carr2010,Slizovskiy2014}. Whether this possibility applies to our three-dimensional system is beyond the scope of this work.

Note that a direct quantitative comparison between our results and the measurements of Aikawa {\it{et al.}} is not meaningful, since they consider dipolar fermions in a trap and argue that first-order interaction corrections due to the time-of-flight expansion cannot be ignored. Taking these effects into account they find good agreement between their Hartree-Fock calculation and the experimental data. Given the smallness of the second-order corrections obtained in our work, it is not surprising that a first-order calculation is sufficient to explain the measurements.
\begin{figure}
\includegraphics[width=\linewidth]{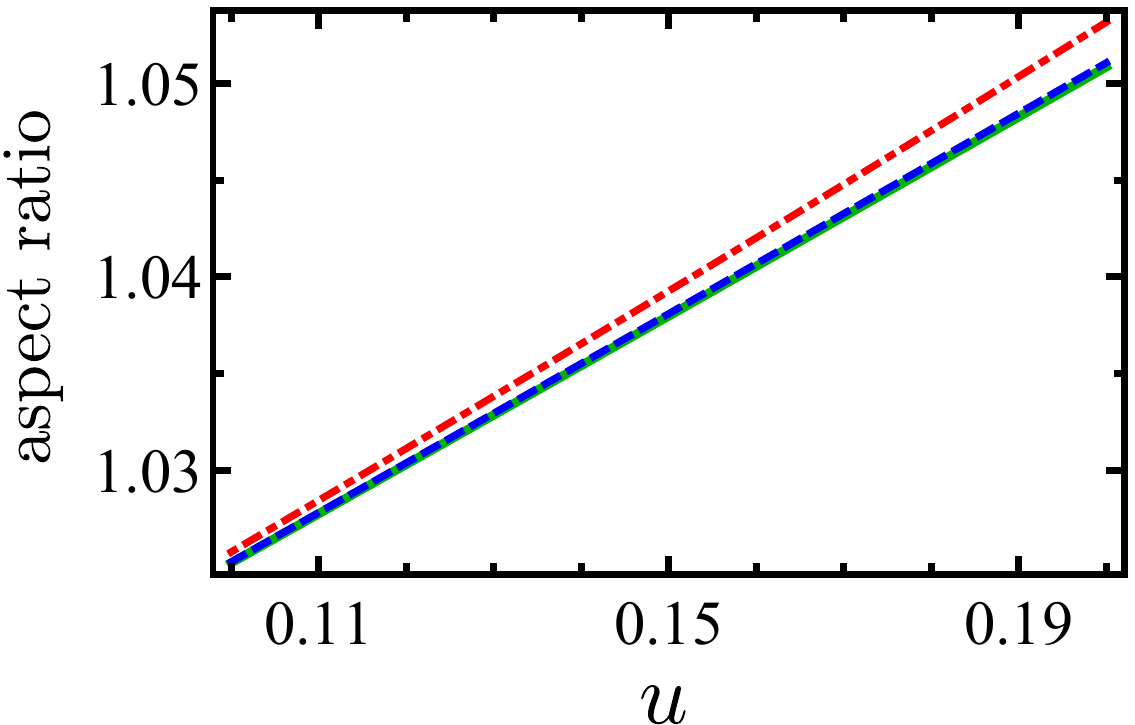}
\caption{
(Color online)
Aspect ratio $k_F (0) / k_F (\frac{\pi}{2})$ of the deformed Fermi surface
for values of $u$ which have recently been reached 
experimentally~\protect\cite{aikawa2014observation}. 
The lines correspond to our results to first order (solid green), 
to second order in the self-consistent Hartree-Fock approximation (dashed blue), and to
full second order (dot-dashed red) in the interaction. While the second-order corrections
taken into account in the self-consistent Hartree-Fock approximation
have practically no effect, the full second-order calculation
changes the first-order result for the aspect ratio by about one per mille which is
of the same order of magnitude as the experimental uncertainty~\protect\cite{aikawa2014observation}.
}
\label{fig:aspect_ratio}
\end{figure}

\subsection{Bulk modulus and instability of the normal state}
\label{subsec:stab_limit}

Combining Eqs.~(\ref{eq:def_r_factor}) and (\ref{eq:r_factor_result}) we find
the renormalized chemical potential to second order in the interaction,
 \begin{equation}
 \mu = \gamma^2 E_{F0} = \frac{k_{F0}^2}{2m} [ 1 - 0.21 u^2 + {\cal{O}} ( u^3 ) ].
 \end{equation}
Using the fact that the density is related to the
bare Fermi momentum as $n = (k_{F0})^3 / (6 \pi^2)$ we obtain
the bulk modulus to second order in the interaction
\begin{equation}
K =     n^2 \biggl(\frac{\partial \mu}{\partial n}\biggr)_{V,T} =      \frac{2}{3} n E_{F0} \left[ 1 - 0.42 u^2 + \mathcal{O}(u^3) \right].
 \label{eq:K2}
\end{equation}
The fact that  the second-order interaction correction to the bulk modulus
is negative suggests
that for sufficiently large values of the interaction the bulk modulus
vanishes and the normal Fermi liquid state becomes unstable in the density-density channel.
Indeed, if we use our second-order result (\ref{eq:K2}) to estimate the critical
interaction strength $u_c$ where $K ( u_c ) =0$ 
 we obtain $u_c \approx 1.55$. This is significantly lower than the estimate based 
on the second-order Hartree-Fock result $u^{\rm HF}_c = 3 \sqrt{10/7} \approx 3.6$ 
[where we neglect the 
Feynman diagrams in Fig.~\ref{fig:diagrams} (c)], while it compares quite well with 
the result $u^{\rm BG}_c \approx 2.1$ obtained by Liu and Yin~\cite{Liu2011a} 
using the 
Brueckner-Goldstone formalism to second order in $u$.
As already mentioned in Sec.~\ref{subsec:introduction} the system may also exhibit other instabilities, e.g., into a biaxial nematic \cite{Fregoso2009} or superfluid \cite{Baranov2002a} phase. While we did not look at this possibility in our work, one should note that these instabilities are in principle allowed and may even precede the density-density instability.

\subsection{Quasiparticle residue and Fermi velocity}
Inserting our results for the frequency derivative 
of the self-energy (which we carried out  analytically before the numerical integration) 
into Eq.~(\ref{eq:qp_res_def}) 
we obtain for the quasiparticle residue 
\begin{eqnarray}
\label{eq:ZExp}
Z_{\bd{k}_F}  & =  & 1 - u^2  \big[ 0.10 + 0.029 P_2 (\cos \alpha)
\nonumber
\\*
& & \hspace{10mm} -
 0.027 P_4 (\cos \alpha) \big] + {\cal{O}} ( u^3 ).
\end{eqnarray}
Similarly, from Eq.~(\ref{eq:vf_def}) we obtain for the modulus of the renormalized Fermi velocity
\begin{eqnarray}
\label{eq:fermiVelocityExp}
\frac{|{\bd{v}}_{\bd{k}_F}|}{v_{F0}} &=& 1 - \frac{u}{12} P_2 (\cos \alpha)  -
u^2 \bigl[ 0.17 + 0.048 P_2 (\cos \alpha)
\nonumber
\\*
&  & \hspace{20mm} - 0.027 P_4 (\cos \alpha) \bigl] + {\cal{O}} ( u^3 ),
 \label{eq:velocity2res}
\end{eqnarray}
where $v_{F0}$ is the bare Fermi velocity. In the upper panel of Fig.~\ref{fig:qp_res} we
show the quasiparticle residue as a function
of the angle $\alpha$ between ${\bd{k}}_F$ and the direction $\hat{\bd{d}}$ and
the dimensionless interaction $u$ in the range
of interactions relevant for the experiment by Aikawa {\it{et al.}}~\cite{aikawa2014observation}.
\begin{figure}
\includegraphics[width=0.95\linewidth]{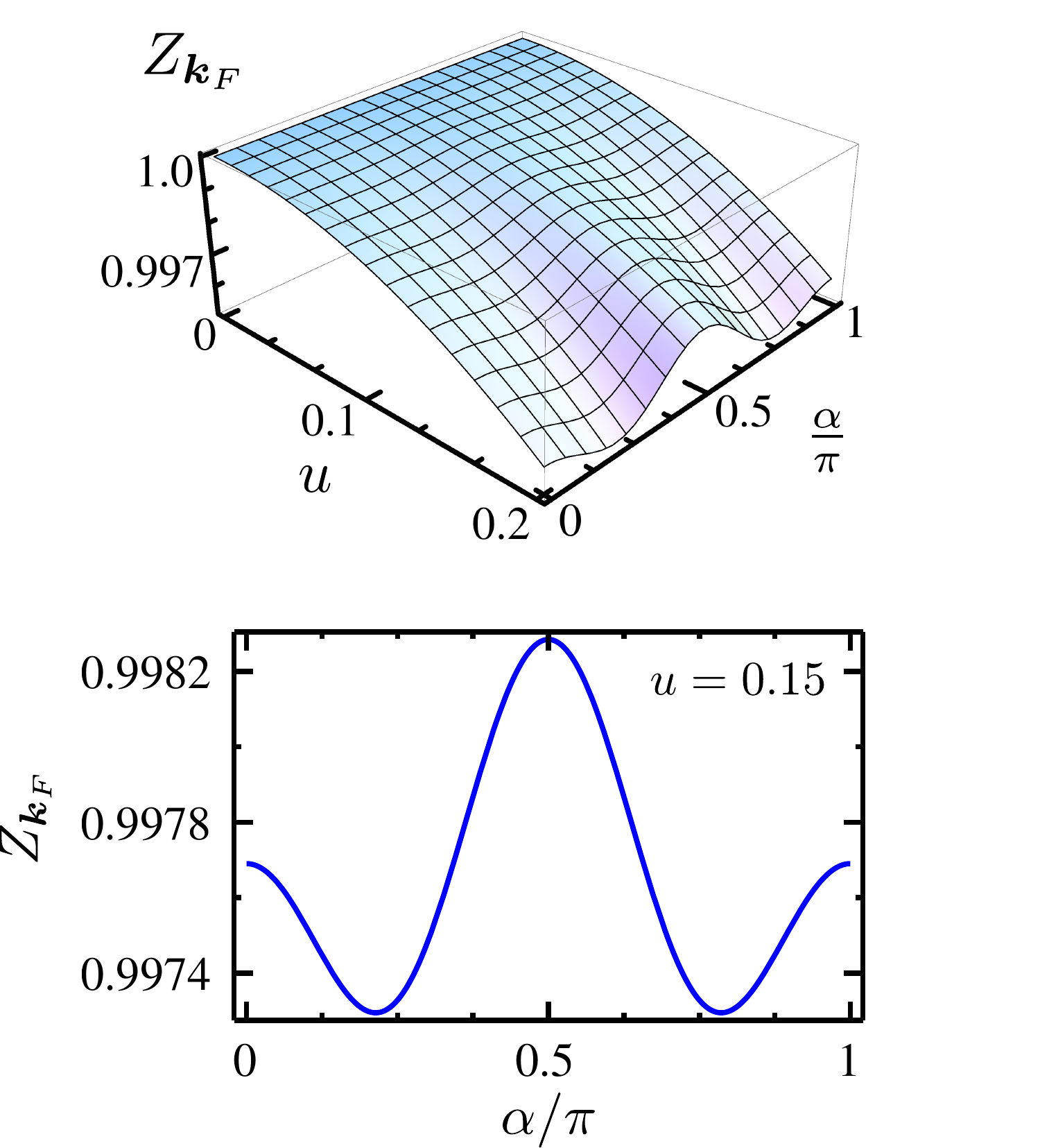}
\caption{
(Color online)
Quasiparticle residue $Z_{\bd{k}_F}$ of dipolar fermions to second order in the interaction; see
Eq.~(\ref{eq:ZExp}).
In the upper panel we show $Z_{\bd{k}_F}$ as a function of the dimensionless
interaction $u  = 8 \pi \nu d^2/3$ and the angle $\alpha$ between $\bd{k}_F$ and
$\hat{\bd{d}}$.  In the lower panel we have fixed $u =0.15$ and show the angular dependence
of the quasiparticle residue.  Note that the value  $u = 0.15$ lies 
in the currently accessible  experimental range~\protect\cite{aikawa2014observation}.
}
\label{fig:qp_res}
\end{figure}
Due to the small value of the interaction, the quasiparticle residue is reduced
only slightly from unity. In the lower panel of  Fig.~\ref{fig:qp_res} 
we show the angular dependence of $Z_{\bd{k}_F}$ for fixed interaction $u =0.15$.
Interestingly, the value of $Z_{\bd{k}_F}$ is smallest
if the angle between
 ${\bd{k}}_F$ and the direction $\hat{\bd{d}}$ is close to $\pi/4$ or $3 \pi /4$.
These local minima are due to the significant 
$P_4 (\cos \alpha)$ component  in the second-order expression for
the quasiparticle residue.

In contrast to the quasiparticle residue, 
the anisotropy of the renormalized Fermi velocity
is for small interactions completely dominated by the 
first-order correction proportional to $P_2 ( \cos \alpha )$.
The renormalized Fermi velocity shown in Fig.~\ref{fig:vf} 
is therefore  directly related to the angular
dependence of the interaction.
\begin{figure}[t]
\includegraphics[width=0.95\linewidth]{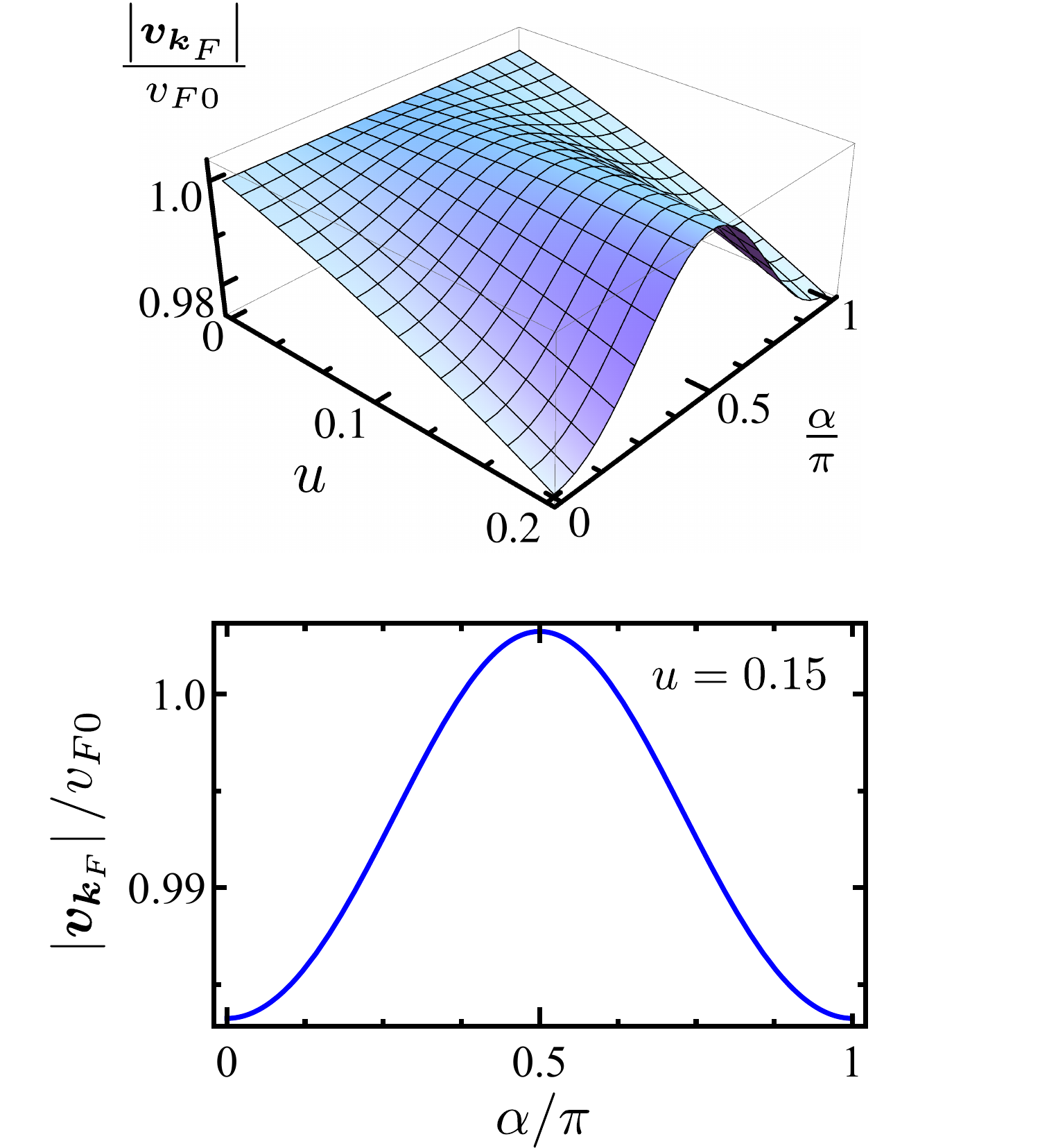}
\caption{
(Color online)
Modulus of the renormalized Fermi velocity ${\bd{v}}_{\bd{k}_F}$ in units of the bare Fermi velocity to second order in
the dimensionless interaction $u$; see Eq.~(\ref{eq:velocity2res}).
 The upper picture shows the behaviour for different interaction strengths, 
while the lower picture shows the velocity for fixed $u = 0.15$ as a function
of the angle $\alpha$ between $\bd{k}_F$ and $\hat{\bd{d}}$.
}
\label{fig:vf}
\end{figure}
If we extrapolate our perturbative result
(\ref{eq:fermiVelocityExp}) for the renormalized Fermi velocity to 
large values of the interaction, we find that
 $|{\bd{v}}_{\bd{k}_F}|$ can become negative for $u \geq 2.09$.
However, from a similar extrapolation of the perturbative expression
for the bulk modulus in Sec.~\ref{subsec:stab_limit} we have found
that the Fermi liquid phase is unstable for $u \geq 1.55$, so that
in the regime where the Fermi liquid phase is stable the
Fermi velocity is always positive.

\subsection{Spectral function}

Finally, let us present our results for the single-particle
spectral function $\rho ( \bd{k} , \omega )$,
which can be obtained by substituting
our numerical results for the real and imaginary parts of the
retarded second-order self-energy into Eq.~(\ref{eq:spectral_def}).
In Fig.~\ref{fig:spectral_functions} we show the spectral function
$\rho (\bd{k}, \omega)$  for wavevectors $\bd{k}$ of the form $\bd{k} =  x \bd{k}_F$,
where the factor $x$ is close to unity.
The spectral line shapes can be very well described by Lorentzians whose width
shrinks to zero as we approach the Fermi surface, as expected
for a Fermi liquid.
Interestingly, for momenta above the Fermi surface the width of the spectral line shape 
(which reflects  the damping of the quasiparticles)
exhibits a rather strong dependence  on the angle $\alpha$ between $\bd{k}_F$ and $\hat{\bd{d}}$,
while for momenta below the Fermi surface the dependence on $\alpha$ is much weaker.
\begin{figure*}[t]
\includegraphics[width=0.85\linewidth]{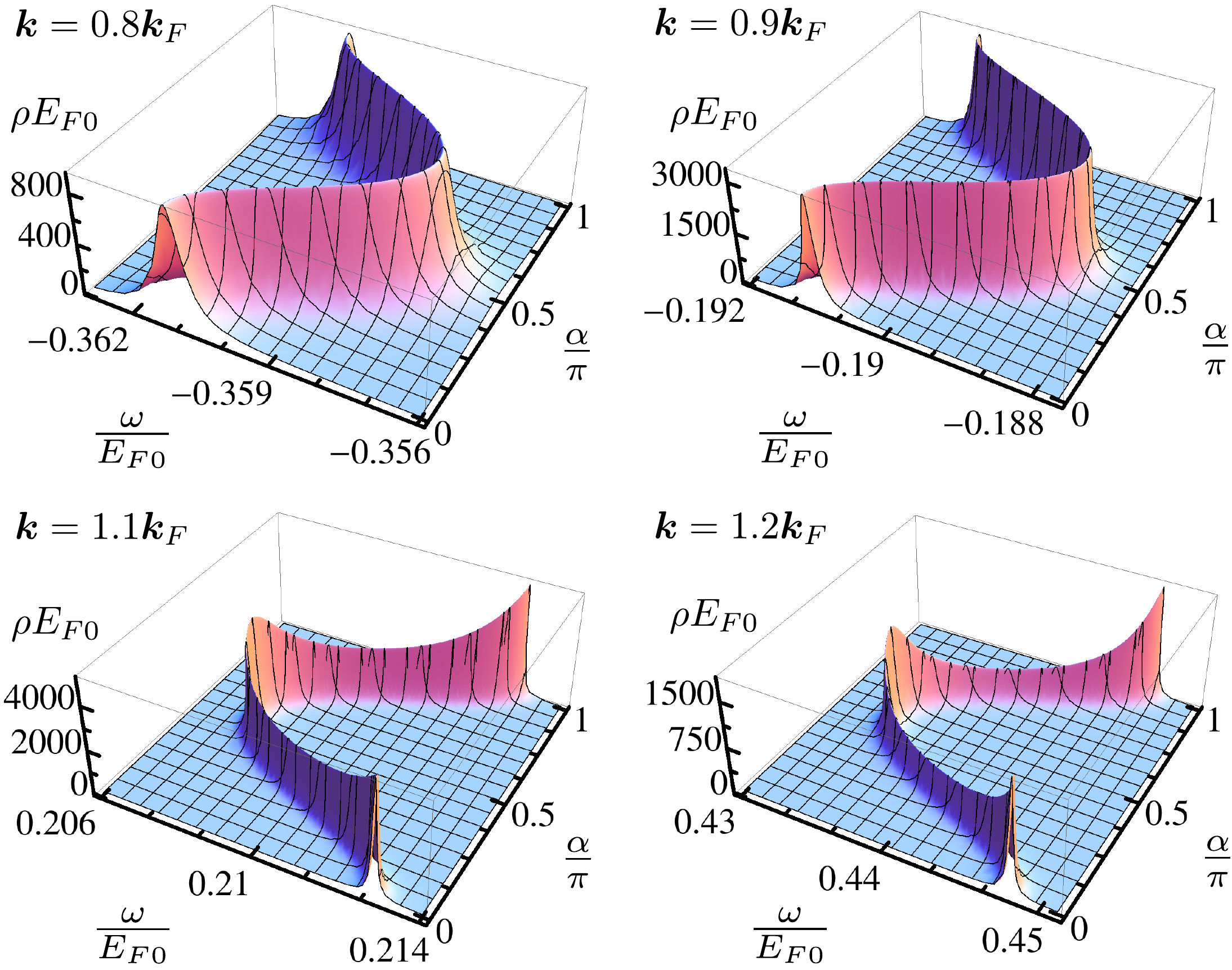}
\caption{
(Color online)
Spectral function $\rho (\bd{k}, \omega)$ for $u = 0.15$ obtained by inserting
our numerical results of the second-order self-energy into
Eq.~(\ref{eq:spectral_def}). 
The upper pictures show the spectral function for excitations 
with wavevectors below the renormalized Fermi surface, 
while the lower pictures give the spectrum for excitations with wavevectors 
above the renormalized Fermi surface. Note that in the latter case the spectral line shape
exhibits a much stronger angular dependence.
}
\label{fig:spectral_functions}
\end{figure*}

\section{Summary and conclusions}
Motivated by a recent experiment
by Aikawa {\it{et al.}}~\cite{aikawa2014observation}
who determined  the Fermi surface of 
a system of $^{167}\text{Er}$ atoms
via time-of-flight measurements, 
we have calculated the self-energy $\Sigma (\bd{k}, \omega)$ of dipolar fermions
in three dimensions to second order in the dipole-dipole interaction.
From this we have inferred the
deformation of the Fermi surface, the quasiparticle residue,
the Fermi velocity, and the spectral function. 
We have shown that the second-order corrections give rise to
a larger elongation of the Fermi surface  and a stronger anisotropy of the Fermi velocity
than the Hartree-Fock approximation.
However,
in the experimentally relevant range of interactions
the second-order corrections are quite small, so that the
Hartree-Fock approximation yields  already quantitatively accurate results.
On the other hand, if in the future it should be possible to 
realize dipolar gases where the effective dimensionless interaction $u= 8 \pi \nu d^2/3$
is of the order of unity, then second-order interaction effects become substantial.
In particular, the angular dependence of the quasiparticle residue $Z_{\bd{k}_F }$ exhibits a 
significant component proportional to $P_4 ( \cos \alpha )$ which can be directly
related to the second-order self-energy.

From our second-order result for the
renormalized  chemical potential we have also estimated
the   critical  interaction strength 
for a collapse instability where the Fermi liquid phase becomes unstable.
Our estimate  $u_c \approx 1.55$ is somewhat smaller than the
result $u_c^{\rm BG} \approx 2.1$ obtained by Liu and Yin \cite{Liu2011a} using the Brueckner-Goldstone 
formalism to second order in $u$. 
Note that the critical $u_c$ is an order of magnitude larger
than the currently experimentally realizable interaction strength.
However, given the rapid experimental progress in the field of ultracold
quantum gases, we hope that larger values of the dimensionless
interaction will be realized in the next few years.
In general, we find that
the interaction corrections to the quasiparticle properties 
 are rather small for all values of $u$ up to the critical interaction $u_c$ where the Fermi liquid exhibits an instability,
so that we conclude that the normal phase of dipolar fermions
can be viewed as a weakly interacting Fermi liquid.

\section*{ACKNOWLEDGMENTS}
We acknowledge financial support by the DFG via FOR 723.




\bibliographystyle{apsrev4-1}
\bibliography{Dipolar_Fermions}

\begin{thebibliography}{67}%
\makeatletter
\providecommand \@ifxundefined [1]{%
 \@ifx{#1\undefined}
}%
\providecommand \@ifnum [1]{%
 \ifnum #1\expandafter \@firstoftwo
 \else \expandafter \@secondoftwo
 \fi
}%
\providecommand \@ifx [1]{%
 \ifx #1\expandafter \@firstoftwo
 \else \expandafter \@secondoftwo
 \fi
}%
\providecommand \natexlab [1]{#1}%
\providecommand \enquote  [1]{``#1''}%
\providecommand \bibnamefont  [1]{#1}%
\providecommand \bibfnamefont [1]{#1}%
\providecommand \citenamefont [1]{#1}%
\providecommand \href@noop [0]{\@secondoftwo}%
\providecommand \href [0]{\begingroup \@sanitize@url \@href}%
\providecommand \@href[1]{\@@startlink{#1}\@@href}%
\providecommand \@@href[1]{\endgroup#1\@@endlink}%
\providecommand \@sanitize@url [0]{\catcode `\\12\catcode `\$12\catcode
  `\&12\catcode `\#12\catcode `\^12\catcode `\_12\catcode `\%12\relax}%
\providecommand \@@startlink[1]{}%
\providecommand \@@endlink[0]{}%
\providecommand \url  [0]{\begingroup\@sanitize@url \@url }%
\providecommand \@url [1]{\endgroup\@href {#1}{\urlprefix }}%
\providecommand \urlprefix  [0]{URL }%
\providecommand \Eprint [0]{\href }%
\providecommand \doibase [0]{http://dx.doi.org/}%
\providecommand \selectlanguage [0]{\@gobble}%
\providecommand \bibinfo  [0]{\@secondoftwo}%
\providecommand \bibfield  [0]{\@secondoftwo}%
\providecommand \translation [1]{[#1]}%
\providecommand \BibitemOpen [0]{}%
\providecommand \bibitemStop [0]{}%
\providecommand \bibitemNoStop [0]{.\EOS\space}%
\providecommand \EOS [0]{\spacefactor3000\relax}%
\providecommand \BibitemShut  [1]{\csname bibitem#1\endcsname}%
\let\auto@bib@innerbib\@empty
\bibitem [{\citenamefont {Chicireanu}\ \emph {et~al.}(2006)\citenamefont
  {Chicireanu}, \citenamefont {Pouderous}, \citenamefont {Barb{\'e}},
  \citenamefont {Laburthe-Tolra}, \citenamefont {Mar{\'e}chal}, \citenamefont
  {Vernac}, \citenamefont {Keller},\ and\ \citenamefont
  {Gorceix}}]{chicireanu2006simultaneous}%
  \BibitemOpen
  \bibfield  {author} {\bibinfo {author} {\bibfnamefont {R.}~\bibnamefont
  {Chicireanu}}, \bibinfo {author} {\bibfnamefont {A.}~\bibnamefont
  {Pouderous}}, \bibinfo {author} {\bibfnamefont {R.}~\bibnamefont
  {Barb{\'e}}}, \bibinfo {author} {\bibfnamefont {B.}~\bibnamefont
  {Laburthe-Tolra}}, \bibinfo {author} {\bibfnamefont {E.}~\bibnamefont
  {Mar{\'e}chal}}, \bibinfo {author} {\bibfnamefont {L.}~\bibnamefont
  {Vernac}}, \bibinfo {author} {\bibfnamefont {J.-C.}\ \bibnamefont {Keller}},
  \ and\ \bibinfo {author} {\bibfnamefont {O.}~\bibnamefont {Gorceix}},\
  }\href@noop {} {\bibfield  {journal} {\bibinfo  {journal} {Phys. Rev. A}\
  }\textbf {\bibinfo {volume} {73}},\ \bibinfo {pages} {053406} (\bibinfo
  {year} {2006})}\BibitemShut {NoStop}%
\bibitem [{\citenamefont {Lu}\ \emph {et~al.}(2012)\citenamefont {Lu},
  \citenamefont {Burdick},\ and\ \citenamefont {Lev}}]{Lu2012}%
  \BibitemOpen
  \bibfield  {author} {\bibinfo {author} {\bibfnamefont {M.}~\bibnamefont
  {Lu}}, \bibinfo {author} {\bibfnamefont {N.~Q.}\ \bibnamefont {Burdick}}, \
  and\ \bibinfo {author} {\bibfnamefont {B.~L.}\ \bibnamefont {Lev}},\
  }\href@noop {} {\bibfield  {journal} {\bibinfo  {journal} {Phys. Rev. Lett.}\
  }\textbf {\bibinfo {volume} {108}},\ \bibinfo {pages} {215301} (\bibinfo
  {year} {2012})}\BibitemShut {NoStop}%
\bibitem [{\citenamefont {Burdick}\ \emph {et~al.}(2015)\citenamefont
  {Burdick}, \citenamefont {Baumann}, \citenamefont {Tang}, \citenamefont
  {Lu},\ and\ \citenamefont {Lev}}]{Burdick2014}%
  \BibitemOpen
  \bibfield  {author} {\bibinfo {author} {\bibfnamefont {N.~Q.}\ \bibnamefont
  {Burdick}}, \bibinfo {author} {\bibfnamefont {K.}~\bibnamefont {Baumann}},
  \bibinfo {author} {\bibfnamefont {Y.}~\bibnamefont {Tang}}, \bibinfo {author}
  {\bibfnamefont {M.}~\bibnamefont {Lu}}, \ and\ \bibinfo {author}
  {\bibfnamefont {B.~L.}\ \bibnamefont {Lev}},\ }\href@noop {} {\bibfield
  {journal} {\bibinfo  {journal} {Phys. Rev. Lett.}\ }\textbf {\bibinfo
  {volume} {114}},\ \bibinfo {pages} {023201} (\bibinfo {year}
  {2015})}\BibitemShut {NoStop}%
\bibitem [{\citenamefont {Aikawa}\ \emph
  {et~al.}(2014{\natexlab{a}})\citenamefont {Aikawa}, \citenamefont {Frisch},
  \citenamefont {Mark}, \citenamefont {Baier}, \citenamefont {Grimm},\ and\
  \citenamefont {Ferlaino}}]{Aikawa2014}%
  \BibitemOpen
  \bibfield  {author} {\bibinfo {author} {\bibfnamefont {K.}~\bibnamefont
  {Aikawa}}, \bibinfo {author} {\bibfnamefont {A.}~\bibnamefont {Frisch}},
  \bibinfo {author} {\bibfnamefont {M.}~\bibnamefont {Mark}}, \bibinfo {author}
  {\bibfnamefont {S.}~\bibnamefont {Baier}}, \bibinfo {author} {\bibfnamefont
  {R.}~\bibnamefont {Grimm}}, \ and\ \bibinfo {author} {\bibfnamefont
  {F.}~\bibnamefont {Ferlaino}},\ }\href@noop {} {\bibfield  {journal}
  {\bibinfo  {journal} {Phys. Rev. Lett.}\ }\textbf {\bibinfo {volume} {112}},\
  \bibinfo {pages} {010404} (\bibinfo {year} {2014}{\natexlab{a}})}\BibitemShut
  {NoStop}%
\bibitem [{\citenamefont {Aikawa}\ \emph
  {et~al.}(2014{\natexlab{b}})\citenamefont {Aikawa}, \citenamefont {Baier},
  \citenamefont {Frisch}, \citenamefont {Mark}, \citenamefont {Ravensbergen},\
  and\ \citenamefont {Ferlaino}}]{aikawa2014observation}%
  \BibitemOpen
  \bibfield  {author} {\bibinfo {author} {\bibfnamefont {K.}~\bibnamefont
  {Aikawa}}, \bibinfo {author} {\bibfnamefont {S.}~\bibnamefont {Baier}},
  \bibinfo {author} {\bibfnamefont {A.}~\bibnamefont {Frisch}}, \bibinfo
  {author} {\bibfnamefont {M.}~\bibnamefont {Mark}}, \bibinfo {author}
  {\bibfnamefont {C.}~\bibnamefont {Ravensbergen}}, \ and\ \bibinfo {author}
  {\bibfnamefont {F.}~\bibnamefont {Ferlaino}},\ }\href@noop {} {\bibfield
  {journal} {\bibinfo  {journal} {Science}\ }\textbf {\bibinfo {volume}
  {345}},\ \bibinfo {pages} {1484} (\bibinfo {year}
  {2014}{\natexlab{b}})}\BibitemShut {NoStop}%
\bibitem [{\citenamefont {Sage}\ \emph {et~al.}(2005)\citenamefont {Sage},
  \citenamefont {Sainis}, \citenamefont {Bergeman},\ and\ \citenamefont
  {DeMille}}]{Sage2005}%
  \BibitemOpen
  \bibfield  {author} {\bibinfo {author} {\bibfnamefont {J.~M.}\ \bibnamefont
  {Sage}}, \bibinfo {author} {\bibfnamefont {S.}~\bibnamefont {Sainis}},
  \bibinfo {author} {\bibfnamefont {T.}~\bibnamefont {Bergeman}}, \ and\
  \bibinfo {author} {\bibfnamefont {D.}~\bibnamefont {DeMille}},\ }\href@noop
  {} {\bibfield  {journal} {\bibinfo  {journal} {Phys. Rev. Lett.}\ }\textbf
  {\bibinfo {volume} {94}},\ \bibinfo {pages} {203001} (\bibinfo {year}
  {2005})}\BibitemShut {NoStop}%
\bibitem [{\citenamefont {Deiglmayr}\ \emph {et~al.}(2008)\citenamefont
  {Deiglmayr}, \citenamefont {Grochola}, \citenamefont {Repp}, \citenamefont
  {M{\"o}rtlbauer}, \citenamefont {Gl{\"u}ck}, \citenamefont {Lange},
  \citenamefont {Dulieu}, \citenamefont {Wester},\ and\ \citenamefont
  {Weidem{\"u}ller}}]{Deiglmayr2008}%
  \BibitemOpen
  \bibfield  {author} {\bibinfo {author} {\bibfnamefont {J.}~\bibnamefont
  {Deiglmayr}}, \bibinfo {author} {\bibfnamefont {A.}~\bibnamefont {Grochola}},
  \bibinfo {author} {\bibfnamefont {M.}~\bibnamefont {Repp}}, \bibinfo {author}
  {\bibfnamefont {K.}~\bibnamefont {M{\"o}rtlbauer}}, \bibinfo {author}
  {\bibfnamefont {C.}~\bibnamefont {Gl{\"u}ck}}, \bibinfo {author}
  {\bibfnamefont {J.}~\bibnamefont {Lange}}, \bibinfo {author} {\bibfnamefont
  {O.}~\bibnamefont {Dulieu}}, \bibinfo {author} {\bibfnamefont
  {R.}~\bibnamefont {Wester}}, \ and\ \bibinfo {author} {\bibfnamefont
  {M.}~\bibnamefont {Weidem{\"u}ller}},\ }\href@noop {} {\bibfield  {journal}
  {\bibinfo  {journal} {Phys. Rev. Lett.}\ }\textbf {\bibinfo {volume} {101}},\
  \bibinfo {pages} {133004} (\bibinfo {year} {2008})}\BibitemShut {NoStop}%
\bibitem [{\citenamefont {Deiglmayr}\ \emph {et~al.}(2010)\citenamefont
  {Deiglmayr}, \citenamefont {Grochola}, \citenamefont {Repp}, \citenamefont
  {Dulieu}, \citenamefont {Wester},\ and\ \citenamefont
  {Weidem{\"u}ller}}]{Deiglmayr2010}%
  \BibitemOpen
  \bibfield  {author} {\bibinfo {author} {\bibfnamefont {J.}~\bibnamefont
  {Deiglmayr}}, \bibinfo {author} {\bibfnamefont {A.}~\bibnamefont {Grochola}},
  \bibinfo {author} {\bibfnamefont {M.}~\bibnamefont {Repp}}, \bibinfo {author}
  {\bibfnamefont {O.}~\bibnamefont {Dulieu}}, \bibinfo {author} {\bibfnamefont
  {R.}~\bibnamefont {Wester}}, \ and\ \bibinfo {author} {\bibfnamefont
  {M.}~\bibnamefont {Weidem{\"u}ller}},\ }\href@noop {} {\bibfield  {journal}
  {\bibinfo  {journal} {Phys. Rev. A}\ }\textbf {\bibinfo {volume} {82}},\
  \bibinfo {pages} {032503} (\bibinfo {year} {2010})}\BibitemShut {NoStop}%
\bibitem [{\citenamefont {Repp}\ \emph {et~al.}(2013)\citenamefont {Repp},
  \citenamefont {Pires}, \citenamefont {Ulmanis}, \citenamefont {Heck},
  \citenamefont {Kuhnle}, \citenamefont {Weidem{\"u}ller},\ and\ \citenamefont
  {Tiemann}}]{Repp2013}%
  \BibitemOpen
  \bibfield  {author} {\bibinfo {author} {\bibfnamefont {M.}~\bibnamefont
  {Repp}}, \bibinfo {author} {\bibfnamefont {R.}~\bibnamefont {Pires}},
  \bibinfo {author} {\bibfnamefont {J.}~\bibnamefont {Ulmanis}}, \bibinfo
  {author} {\bibfnamefont {R.}~\bibnamefont {Heck}}, \bibinfo {author}
  {\bibfnamefont {E.~D.}\ \bibnamefont {Kuhnle}}, \bibinfo {author}
  {\bibfnamefont {M.}~\bibnamefont {Weidem{\"u}ller}}, \ and\ \bibinfo {author}
  {\bibfnamefont {E.}~\bibnamefont {Tiemann}},\ }\href@noop {} {\bibfield
  {journal} {\bibinfo  {journal} {Phys. Rev. A}\ }\textbf {\bibinfo {volume}
  {87}},\ \bibinfo {pages} {010701} (\bibinfo {year} {2013})}\BibitemShut
  {NoStop}%
\bibitem [{\citenamefont {Heo}\ \emph {et~al.}(2012)\citenamefont {Heo},
  \citenamefont {Wang}, \citenamefont {Christensen}, \citenamefont {Rvachov},
  \citenamefont {Cotta}, \citenamefont {Choi}, \citenamefont {Lee},\ and\
  \citenamefont {Ketterle}}]{Heo2012}%
  \BibitemOpen
  \bibfield  {author} {\bibinfo {author} {\bibfnamefont {M.-S.}\ \bibnamefont
  {Heo}}, \bibinfo {author} {\bibfnamefont {T.~T.}\ \bibnamefont {Wang}},
  \bibinfo {author} {\bibfnamefont {C.~A.}\ \bibnamefont {Christensen}},
  \bibinfo {author} {\bibfnamefont {T.~M.}\ \bibnamefont {Rvachov}}, \bibinfo
  {author} {\bibfnamefont {D.~A.}\ \bibnamefont {Cotta}}, \bibinfo {author}
  {\bibfnamefont {J.-H.}\ \bibnamefont {Choi}}, \bibinfo {author}
  {\bibfnamefont {Y.-R.}\ \bibnamefont {Lee}}, \ and\ \bibinfo {author}
  {\bibfnamefont {W.}~\bibnamefont {Ketterle}},\ }\href@noop {} {\bibfield
  {journal} {\bibinfo  {journal} {Phys. Rev. A}\ }\textbf {\bibinfo {volume}
  {86}},\ \bibinfo {pages} {021602} (\bibinfo {year} {2012})}\BibitemShut
  {NoStop}%
\bibitem [{\citenamefont {Wu}\ \emph {et~al.}(2012)\citenamefont {Wu},
  \citenamefont {Park}, \citenamefont {Ahmadi}, \citenamefont {Will},\ and\
  \citenamefont {Zwierlein}}]{Wu2012}%
  \BibitemOpen
  \bibfield  {author} {\bibinfo {author} {\bibfnamefont {C.-H.}\ \bibnamefont
  {Wu}}, \bibinfo {author} {\bibfnamefont {J.~W.}\ \bibnamefont {Park}},
  \bibinfo {author} {\bibfnamefont {P.}~\bibnamefont {Ahmadi}}, \bibinfo
  {author} {\bibfnamefont {S.}~\bibnamefont {Will}}, \ and\ \bibinfo {author}
  {\bibfnamefont {M.~W.}\ \bibnamefont {Zwierlein}},\ }\href@noop {} {\bibfield
   {journal} {\bibinfo  {journal} {Phys. Rev. Lett.}\ }\textbf {\bibinfo
  {volume} {109}},\ \bibinfo {pages} {085301} (\bibinfo {year}
  {2012})}\BibitemShut {NoStop}%
\bibitem [{\citenamefont {Ospelkaus}\ \emph {et~al.}(2008)\citenamefont
  {Ospelkaus}, \citenamefont {Pe'er}, \citenamefont {Ni}, \citenamefont
  {Zirbel}, \citenamefont {Neyenhuis}, \citenamefont {Kotochigova},
  \citenamefont {Julienne}, \citenamefont {Ye},\ and\ \citenamefont
  {Jin}}]{Ospelkaus2008}%
  \BibitemOpen
  \bibfield  {author} {\bibinfo {author} {\bibfnamefont {S.}~\bibnamefont
  {Ospelkaus}}, \bibinfo {author} {\bibfnamefont {A.}~\bibnamefont {Pe'er}},
  \bibinfo {author} {\bibfnamefont {K.-K.}\ \bibnamefont {Ni}}, \bibinfo
  {author} {\bibfnamefont {J.~J.}\ \bibnamefont {Zirbel}}, \bibinfo {author}
  {\bibfnamefont {B.}~\bibnamefont {Neyenhuis}}, \bibinfo {author}
  {\bibfnamefont {S.}~\bibnamefont {Kotochigova}}, \bibinfo {author}
  {\bibfnamefont {P.~S.}\ \bibnamefont {Julienne}}, \bibinfo {author}
  {\bibfnamefont {J.}~\bibnamefont {Ye}}, \ and\ \bibinfo {author}
  {\bibfnamefont {D.~S.}\ \bibnamefont {Jin}},\ }\href@noop {} {\bibfield
  {journal} {\bibinfo  {journal} {Nature Phys.}\ }\textbf {\bibinfo {volume}
  {4}},\ \bibinfo {pages} {622} (\bibinfo {year} {2008})}\BibitemShut {NoStop}%
\bibitem [{\citenamefont {Ni}\ \emph {et~al.}(2008)\citenamefont {Ni},
  \citenamefont {Ospelkaus}, \citenamefont {de~Miranda}, \citenamefont {Pe'er},
  \citenamefont {Neyenhuis}, \citenamefont {Zirbel}, \citenamefont
  {Kotochigova}, \citenamefont {Julienne}, \citenamefont {Jin},\ and\
  \citenamefont {Ye}}]{Ni2008}%
  \BibitemOpen
  \bibfield  {author} {\bibinfo {author} {\bibfnamefont {K.-K.}\ \bibnamefont
  {Ni}}, \bibinfo {author} {\bibfnamefont {S.}~\bibnamefont {Ospelkaus}},
  \bibinfo {author} {\bibfnamefont {M.~H.~G.}\ \bibnamefont {de~Miranda}},
  \bibinfo {author} {\bibfnamefont {A.}~\bibnamefont {Pe'er}}, \bibinfo
  {author} {\bibfnamefont {B.}~\bibnamefont {Neyenhuis}}, \bibinfo {author}
  {\bibfnamefont {J.~J.}\ \bibnamefont {Zirbel}}, \bibinfo {author}
  {\bibfnamefont {S.}~\bibnamefont {Kotochigova}}, \bibinfo {author}
  {\bibfnamefont {P.~S.}\ \bibnamefont {Julienne}}, \bibinfo {author}
  {\bibfnamefont {D.~S.}\ \bibnamefont {Jin}}, \ and\ \bibinfo {author}
  {\bibfnamefont {J.}~\bibnamefont {Ye}},\ }\href@noop {} {\bibfield  {journal}
  {\bibinfo  {journal} {Science}\ }\textbf {\bibinfo {volume} {322}},\ \bibinfo
  {pages} {231} (\bibinfo {year} {2008})}\BibitemShut {NoStop}%
\bibitem [{\citenamefont {Ospelkaus}\ \emph {et~al.}(2009)\citenamefont
  {Ospelkaus}, \citenamefont {Ni}, \citenamefont {de~Miranda}, \citenamefont
  {Neyenhuis}, \citenamefont {Wang}, \citenamefont {Kotochigova}, \citenamefont
  {Julienne}, \citenamefont {Jin},\ and\ \citenamefont
  {Ye}}]{ade2009ultracold}%
  \BibitemOpen
  \bibfield  {author} {\bibinfo {author} {\bibfnamefont {S.}~\bibnamefont
  {Ospelkaus}}, \bibinfo {author} {\bibfnamefont {K.-K.}\ \bibnamefont {Ni}},
  \bibinfo {author} {\bibfnamefont {M.~H.~G.}\ \bibnamefont {de~Miranda}},
  \bibinfo {author} {\bibfnamefont {B.}~\bibnamefont {Neyenhuis}}, \bibinfo
  {author} {\bibfnamefont {D.}~\bibnamefont {Wang}}, \bibinfo {author}
  {\bibfnamefont {S.}~\bibnamefont {Kotochigova}}, \bibinfo {author}
  {\bibfnamefont {P.~S.}\ \bibnamefont {Julienne}}, \bibinfo {author}
  {\bibfnamefont {D.~S.}\ \bibnamefont {Jin}}, \ and\ \bibinfo {author}
  {\bibfnamefont {J.}~\bibnamefont {Ye}},\ }\href@noop {} {\bibfield  {journal}
  {\bibinfo  {journal} {Farad. Discuss.}\ }\textbf {\bibinfo {volume} {142}},\
  \bibinfo {pages} {351} (\bibinfo {year} {2009})}\BibitemShut {NoStop}%
\bibitem [{\citenamefont {Ni}\ \emph {et~al.}(2010)\citenamefont {Ni},
  \citenamefont {Ospelkaus}, \citenamefont {Wang}, \citenamefont
  {Qu{\'e}m{\'e}ner}, \citenamefont {Neyenhuis}, \citenamefont {de~Miranda},
  \citenamefont {Bohn}, \citenamefont {Ye},\ and\ \citenamefont
  {Jin}}]{ni2010dipolar}%
  \BibitemOpen
  \bibfield  {author} {\bibinfo {author} {\bibfnamefont {K.-K.}\ \bibnamefont
  {Ni}}, \bibinfo {author} {\bibfnamefont {S.}~\bibnamefont {Ospelkaus}},
  \bibinfo {author} {\bibfnamefont {D.}~\bibnamefont {Wang}}, \bibinfo {author}
  {\bibfnamefont {G.}~\bibnamefont {Qu{\'e}m{\'e}ner}}, \bibinfo {author}
  {\bibfnamefont {B.}~\bibnamefont {Neyenhuis}}, \bibinfo {author}
  {\bibfnamefont {M.~H.~G.}\ \bibnamefont {de~Miranda}}, \bibinfo {author}
  {\bibfnamefont {J.~L.}\ \bibnamefont {Bohn}}, \bibinfo {author}
  {\bibfnamefont {J.}~\bibnamefont {Ye}}, \ and\ \bibinfo {author}
  {\bibfnamefont {D.~S.}\ \bibnamefont {Jin}},\ }\href@noop {} {\bibfield
  {journal} {\bibinfo  {journal} {Nature}\ }\textbf {\bibinfo {volume} {464}},\
  \bibinfo {pages} {1324} (\bibinfo {year} {2010})}\BibitemShut {NoStop}%
\bibitem [{\citenamefont {de~Miranda}\ \emph {et~al.}(2011)\citenamefont
  {de~Miranda}, \citenamefont {Chotia}, \citenamefont {Neyenhuis},
  \citenamefont {Wang}, \citenamefont {Qu{\'e}m{\'e}ner}, \citenamefont
  {Ospelkaus}, \citenamefont {Bohn}, \citenamefont {Ye},\ and\ \citenamefont
  {Jin}}]{DeMiranda2011}%
  \BibitemOpen
  \bibfield  {author} {\bibinfo {author} {\bibfnamefont {M.~H.~G.}\
  \bibnamefont {de~Miranda}}, \bibinfo {author} {\bibfnamefont
  {A.}~\bibnamefont {Chotia}}, \bibinfo {author} {\bibfnamefont
  {B.}~\bibnamefont {Neyenhuis}}, \bibinfo {author} {\bibfnamefont
  {D.}~\bibnamefont {Wang}}, \bibinfo {author} {\bibfnamefont {G.}~\bibnamefont
  {Qu{\'e}m{\'e}ner}}, \bibinfo {author} {\bibfnamefont {S.}~\bibnamefont
  {Ospelkaus}}, \bibinfo {author} {\bibfnamefont {J.~L.}\ \bibnamefont {Bohn}},
  \bibinfo {author} {\bibfnamefont {J.}~\bibnamefont {Ye}}, \ and\ \bibinfo
  {author} {\bibfnamefont {D.~S.}\ \bibnamefont {Jin}},\ }\href@noop {}
  {\bibfield  {journal} {\bibinfo  {journal} {Nature Phys.}\ }\textbf {\bibinfo
  {volume} {7}},\ \bibinfo {pages} {502} (\bibinfo {year} {2011})}\BibitemShut
  {NoStop}%
\bibitem [{\citenamefont {Baranov}(2008)}]{Bar08}%
  \BibitemOpen
  \bibfield  {author} {\bibinfo {author} {\bibfnamefont {M.~A.}\ \bibnamefont
  {Baranov}},\ }\href@noop {} {\bibfield  {journal} {\bibinfo  {journal} {Phys.
  Rep.}\ }\textbf {\bibinfo {volume} {464}},\ \bibinfo {pages} {71} (\bibinfo
  {year} {2008})}\BibitemShut {NoStop}%
\bibitem [{\citenamefont {Baranov}\ \emph {et~al.}(2012)\citenamefont
  {Baranov}, \citenamefont {Dalmonte}, \citenamefont {Pupillo},\ and\
  \citenamefont {Zoller}}]{Bar12}%
  \BibitemOpen
  \bibfield  {author} {\bibinfo {author} {\bibfnamefont {M.~A.}\ \bibnamefont
  {Baranov}}, \bibinfo {author} {\bibfnamefont {M.}~\bibnamefont {Dalmonte}},
  \bibinfo {author} {\bibfnamefont {G.}~\bibnamefont {Pupillo}}, \ and\
  \bibinfo {author} {\bibfnamefont {P.}~\bibnamefont {Zoller}},\ }\href@noop {}
  {\bibfield  {journal} {\bibinfo  {journal} {Chem. Rev.}\ }\textbf {\bibinfo
  {volume} {112}},\ \bibinfo {pages} {5012} (\bibinfo {year}
  {2012})}\BibitemShut {NoStop}%
\bibitem [{\citenamefont {Chan}\ \emph {et~al.}(2010)\citenamefont {Chan},
  \citenamefont {Wu}, \citenamefont {Lee},\ and\ \citenamefont
  {Das~Sarma}}]{Cha10}%
  \BibitemOpen
  \bibfield  {author} {\bibinfo {author} {\bibfnamefont {C.-K.}\ \bibnamefont
  {Chan}}, \bibinfo {author} {\bibfnamefont {C.}~\bibnamefont {Wu}}, \bibinfo
  {author} {\bibfnamefont {W.-C.}\ \bibnamefont {Lee}}, \ and\ \bibinfo
  {author} {\bibfnamefont {S.}~\bibnamefont {Das~Sarma}},\ }\href@noop {}
  {\bibfield  {journal} {\bibinfo  {journal} {Phys. Rev. A}\ }\textbf {\bibinfo
  {volume} {81}},\ \bibinfo {pages} {023602} (\bibinfo {year}
  {2010})}\BibitemShut {NoStop}%
\bibitem [{\citenamefont {Yamaguchi}\ \emph {et~al.}(2010)\citenamefont
  {Yamaguchi}, \citenamefont {Sogo}, \citenamefont {Ito},\ and\ \citenamefont
  {Miyakawa}}]{Yamaguchi2010}%
  \BibitemOpen
  \bibfield  {author} {\bibinfo {author} {\bibfnamefont {Y.}~\bibnamefont
  {Yamaguchi}}, \bibinfo {author} {\bibfnamefont {T.}~\bibnamefont {Sogo}},
  \bibinfo {author} {\bibfnamefont {T.}~\bibnamefont {Ito}}, \ and\ \bibinfo
  {author} {\bibfnamefont {T.}~\bibnamefont {Miyakawa}},\ }\href@noop {}
  {\bibfield  {journal} {\bibinfo  {journal} {Phys. Rev. A}\ }\textbf {\bibinfo
  {volume} {82}},\ \bibinfo {pages} {013643} (\bibinfo {year}
  {2010})}\BibitemShut {NoStop}%
\bibitem [{\citenamefont {Lu}\ \emph {et~al.}(2013)\citenamefont {Lu},
  \citenamefont {Matveenko},\ and\ \citenamefont {Shlyapnikov}}]{Lu2013}%
  \BibitemOpen
  \bibfield  {author} {\bibinfo {author} {\bibfnamefont {Z.-K.}\ \bibnamefont
  {Lu}}, \bibinfo {author} {\bibfnamefont {S.~I.}\ \bibnamefont {Matveenko}}, \
  and\ \bibinfo {author} {\bibfnamefont {G.~V.}\ \bibnamefont {Shlyapnikov}},\
  }\href@noop {} {\bibfield  {journal} {\bibinfo  {journal} {Phys. Rev. A}\
  }\textbf {\bibinfo {volume} {88}},\ \bibinfo {pages} {033625} (\bibinfo
  {year} {2013})}\BibitemShut {NoStop}%
\bibitem [{\citenamefont {G{\'o}ral}\ \emph {et~al.}(2001)\citenamefont
  {G{\'o}ral}, \citenamefont {Englert},\ and\ \citenamefont
  {Rza̧{\.z}ewski}}]{Goral2001}%
  \BibitemOpen
  \bibfield  {author} {\bibinfo {author} {\bibfnamefont {K.}~\bibnamefont
  {G{\'o}ral}}, \bibinfo {author} {\bibfnamefont {B.-G.}\ \bibnamefont
  {Englert}}, \ and\ \bibinfo {author} {\bibfnamefont {K.}~\bibnamefont
  {Rza̧{\.z}ewski}},\ }\href@noop {} {\bibfield  {journal} {\bibinfo
  {journal} {Phys. Rev. A}\ }\textbf {\bibinfo {volume} {63}},\ \bibinfo
  {pages} {033606} (\bibinfo {year} {2001})}\BibitemShut {NoStop}%
\bibitem [{\citenamefont {Miyakawa}\ \emph {et~al.}(2008)\citenamefont
  {Miyakawa}, \citenamefont {Sogo},\ and\ \citenamefont {Pu}}]{Miy08}%
  \BibitemOpen
  \bibfield  {author} {\bibinfo {author} {\bibfnamefont {T.}~\bibnamefont
  {Miyakawa}}, \bibinfo {author} {\bibfnamefont {T.}~\bibnamefont {Sogo}}, \
  and\ \bibinfo {author} {\bibfnamefont {H.}~\bibnamefont {Pu}},\ }\href@noop
  {} {\bibfield  {journal} {\bibinfo  {journal} {Phys. Rev. A}\ }\textbf
  {\bibinfo {volume} {77}},\ \bibinfo {pages} {061603} (\bibinfo {year}
  {2008})}\BibitemShut {NoStop}%
\bibitem [{\citenamefont {Fregoso}\ \emph {et~al.}(2009)\citenamefont
  {Fregoso}, \citenamefont {Sun}, \citenamefont {Fradkin},\ and\ \citenamefont
  {Lev}}]{Fregoso2009}%
  \BibitemOpen
  \bibfield  {author} {\bibinfo {author} {\bibfnamefont {B.~M.}\ \bibnamefont
  {Fregoso}}, \bibinfo {author} {\bibfnamefont {K.}~\bibnamefont {Sun}},
  \bibinfo {author} {\bibfnamefont {E.}~\bibnamefont {Fradkin}}, \ and\
  \bibinfo {author} {\bibfnamefont {B.~L.}\ \bibnamefont {Lev}},\ }\href@noop
  {} {\bibfield  {journal} {\bibinfo  {journal} {New J. Phys.}\ }\textbf
  {\bibinfo {volume} {11}},\ \bibinfo {pages} {103003} (\bibinfo {year}
  {2009})}\BibitemShut {NoStop}%
\bibitem [{\citenamefont {Fregoso}\ and\ \citenamefont
  {Fradkin}(2010)}]{Fregoso2010}%
  \BibitemOpen
  \bibfield  {author} {\bibinfo {author} {\bibfnamefont {B.~M.}\ \bibnamefont
  {Fregoso}}\ and\ \bibinfo {author} {\bibfnamefont {E.}~\bibnamefont
  {Fradkin}},\ }\href@noop {} {\bibfield  {journal} {\bibinfo  {journal} {Phys.
  Rev. B}\ }\textbf {\bibinfo {volume} {81}},\ \bibinfo {pages} {214443}
  (\bibinfo {year} {2010})}\BibitemShut {NoStop}%
\bibitem [{\citenamefont {Ronen}\ and\ \citenamefont {Bohn}(2010)}]{Ronen10}%
  \BibitemOpen
  \bibfield  {author} {\bibinfo {author} {\bibfnamefont {S.}~\bibnamefont
  {Ronen}}\ and\ \bibinfo {author} {\bibfnamefont {J.~L.}\ \bibnamefont
  {Bohn}},\ }\href@noop {} {\bibfield  {journal} {\bibinfo  {journal} {Phys.
  Rev. A}\ }\textbf {\bibinfo {volume} {81}},\ \bibinfo {pages} {033601}
  (\bibinfo {year} {2010})}\BibitemShut {NoStop}%
\bibitem [{\citenamefont {Baillie}\ and\ \citenamefont
  {Blakie}(2010)}]{Baillie2010}%
  \BibitemOpen
  \bibfield  {author} {\bibinfo {author} {\bibfnamefont {D.}~\bibnamefont
  {Baillie}}\ and\ \bibinfo {author} {\bibfnamefont {P.~B.}\ \bibnamefont
  {Blakie}},\ }\href@noop {} {\bibfield  {journal} {\bibinfo  {journal} {Phys.
  Rev. A}\ }\textbf {\bibinfo {volume} {82}},\ \bibinfo {pages} {033605}
  (\bibinfo {year} {2010})}\BibitemShut {NoStop}%
\bibitem [{\citenamefont {Baranov}\ \emph {et~al.}(2002)\citenamefont
  {Baranov}, \citenamefont {Mar'enko}, \citenamefont {Rychkov},\ and\
  \citenamefont {Shlyapnikov}}]{Baranov2002a}%
  \BibitemOpen
  \bibfield  {author} {\bibinfo {author} {\bibfnamefont {M.~A.}\ \bibnamefont
  {Baranov}}, \bibinfo {author} {\bibfnamefont {M.~S.}\ \bibnamefont
  {Mar'enko}}, \bibinfo {author} {\bibfnamefont {V.~S.}\ \bibnamefont
  {Rychkov}}, \ and\ \bibinfo {author} {\bibfnamefont {G.~V.}\ \bibnamefont
  {Shlyapnikov}},\ }\href@noop {} {\bibfield  {journal} {\bibinfo  {journal}
  {Phys. Rev. A}\ }\textbf {\bibinfo {volume} {66}},\ \bibinfo {pages} {013606}
  (\bibinfo {year} {2002})}\BibitemShut {NoStop}%
\bibitem [{\citenamefont {Bruun}\ and\ \citenamefont
  {Taylor}(2008)}]{Bruun2008}%
  \BibitemOpen
  \bibfield  {author} {\bibinfo {author} {\bibfnamefont {G.~M.}\ \bibnamefont
  {Bruun}}\ and\ \bibinfo {author} {\bibfnamefont {E.}~\bibnamefont {Taylor}},\
  }\href@noop {} {\bibfield  {journal} {\bibinfo  {journal} {Phys. Rev. Lett.}\
  }\textbf {\bibinfo {volume} {101}},\ \bibinfo {pages} {245301} (\bibinfo
  {year} {2008})}\BibitemShut {NoStop}%
\bibitem [{\citenamefont {Cooper}\ and\ \citenamefont
  {Shlyapnikov}(2009)}]{Cooper2009}%
  \BibitemOpen
  \bibfield  {author} {\bibinfo {author} {\bibfnamefont {N.~R.}\ \bibnamefont
  {Cooper}}\ and\ \bibinfo {author} {\bibfnamefont {G.~V.}\ \bibnamefont
  {Shlyapnikov}},\ }\href@noop {} {\bibfield  {journal} {\bibinfo  {journal}
  {Phys. Rev. Lett.}\ }\textbf {\bibinfo {volume} {103}},\ \bibinfo {pages}
  {155302} (\bibinfo {year} {2009})}\BibitemShut {NoStop}%
\bibitem [{\citenamefont {Zhao}\ \emph {et~al.}(2010)\citenamefont {Zhao},
  \citenamefont {Jiang}, \citenamefont {Liu}, \citenamefont {Liu},
  \citenamefont {Zou},\ and\ \citenamefont {Pu}}]{Zhao2010}%
  \BibitemOpen
  \bibfield  {author} {\bibinfo {author} {\bibfnamefont {C.}~\bibnamefont
  {Zhao}}, \bibinfo {author} {\bibfnamefont {L.}~\bibnamefont {Jiang}},
  \bibinfo {author} {\bibfnamefont {X.}~\bibnamefont {Liu}}, \bibinfo {author}
  {\bibfnamefont {W.~M.}\ \bibnamefont {Liu}}, \bibinfo {author} {\bibfnamefont
  {X.}~\bibnamefont {Zou}}, \ and\ \bibinfo {author} {\bibfnamefont
  {H.}~\bibnamefont {Pu}},\ }\href@noop {} {\bibfield  {journal} {\bibinfo
  {journal} {Phys. Rev. A}\ }\textbf {\bibinfo {volume} {81}},\ \bibinfo
  {pages} {063642} (\bibinfo {year} {2010})}\BibitemShut {NoStop}%
\bibitem [{\citenamefont {Liao}\ and\ \citenamefont {Brand}(2010)}]{Liao2010}%
  \BibitemOpen
  \bibfield  {author} {\bibinfo {author} {\bibfnamefont {R.}~\bibnamefont
  {Liao}}\ and\ \bibinfo {author} {\bibfnamefont {J.}~\bibnamefont {Brand}},\
  }\href@noop {} {\bibfield  {journal} {\bibinfo  {journal} {Phys. Rev. A}\
  }\textbf {\bibinfo {volume} {82}},\ \bibinfo {pages} {063624} (\bibinfo
  {year} {2010})}\BibitemShut {NoStop}%
\bibitem [{\citenamefont {Fregoso}\ and\ \citenamefont
  {Fradkin}(2009)}]{Fregoso2009a}%
  \BibitemOpen
  \bibfield  {author} {\bibinfo {author} {\bibfnamefont {B.~M.}\ \bibnamefont
  {Fregoso}}\ and\ \bibinfo {author} {\bibfnamefont {E.}~\bibnamefont
  {Fradkin}},\ }\href@noop {} {\bibfield  {journal} {\bibinfo  {journal} {Phys.
  Rev. Lett.}\ }\textbf {\bibinfo {volume} {103}},\ \bibinfo {pages} {205301}
  (\bibinfo {year} {2009})}\BibitemShut {NoStop}%
\bibitem [{\citenamefont {Quintanilla}\ \emph {et~al.}(2009)\citenamefont
  {Quintanilla}, \citenamefont {Carr},\ and\ \citenamefont
  {Betouras}}]{Quintanilla2009}%
  \BibitemOpen
  \bibfield  {author} {\bibinfo {author} {\bibfnamefont {J.}~\bibnamefont
  {Quintanilla}}, \bibinfo {author} {\bibfnamefont {S.~T.}\ \bibnamefont
  {Carr}}, \ and\ \bibinfo {author} {\bibfnamefont {J.~J.}\ \bibnamefont
  {Betouras}},\ }\href@noop {} {\bibfield  {journal} {\bibinfo  {journal}
  {Phys. Rev. A}\ }\textbf {\bibinfo {volume} {79}},\ \bibinfo {pages} {031601}
  (\bibinfo {year} {2009})}\BibitemShut {NoStop}%
\bibitem [{\citenamefont {Carr}\ \emph {et~al.}(2009)\citenamefont {Carr},
  \citenamefont {Quintanilla},\ and\ \citenamefont {Betouras}}]{Carr2009}%
  \BibitemOpen
  \bibfield  {author} {\bibinfo {author} {\bibfnamefont {S.~T.}\ \bibnamefont
  {Carr}}, \bibinfo {author} {\bibfnamefont {J.}~\bibnamefont {Quintanilla}}, \
  and\ \bibinfo {author} {\bibfnamefont {J.~J.}\ \bibnamefont {Betouras}},\
  }\href@noop {} {\bibfield  {journal} {\bibinfo  {journal} {Int. J. Mod. Phys.
  B}\ }\textbf {\bibinfo {volume} {23}},\ \bibinfo {pages} {4074} (\bibinfo
  {year} {2009})}\BibitemShut {NoStop}%
\bibitem [{\citenamefont {Maeda}\ \emph {et~al.}(2013)\citenamefont {Maeda},
  \citenamefont {Hatsuda},\ and\ \citenamefont {Baym}}]{Maeda2013}%
  \BibitemOpen
  \bibfield  {author} {\bibinfo {author} {\bibfnamefont {K.}~\bibnamefont
  {Maeda}}, \bibinfo {author} {\bibfnamefont {T.}~\bibnamefont {Hatsuda}}, \
  and\ \bibinfo {author} {\bibfnamefont {G.}~\bibnamefont {Baym}},\ }\href@noop
  {} {\bibfield  {journal} {\bibinfo  {journal} {Phys. Rev. A}\ }\textbf
  {\bibinfo {volume} {87}},\ \bibinfo {pages} {021604} (\bibinfo {year}
  {2013})}\BibitemShut {NoStop}%
\bibitem [{\citenamefont {Sun}\ \emph {et~al.}(2010)\citenamefont {Sun},
  \citenamefont {Wu},\ and\ \citenamefont {Das~Sarma}}]{Sun2010}%
  \BibitemOpen
  \bibfield  {author} {\bibinfo {author} {\bibfnamefont {K.}~\bibnamefont
  {Sun}}, \bibinfo {author} {\bibfnamefont {C.}~\bibnamefont {Wu}}, \ and\
  \bibinfo {author} {\bibfnamefont {S.}~\bibnamefont {Das~Sarma}},\ }\href@noop
  {} {\bibfield  {journal} {\bibinfo  {journal} {Phys. Rev. B}\ }\textbf
  {\bibinfo {volume} {82}},\ \bibinfo {pages} {075105} (\bibinfo {year}
  {2010})}\BibitemShut {NoStop}%
\bibitem [{\citenamefont {Mikelsons}\ and\ \citenamefont
  {Freericks}(2011)}]{Mikelsons2011}%
  \BibitemOpen
  \bibfield  {author} {\bibinfo {author} {\bibfnamefont {K.}~\bibnamefont
  {Mikelsons}}\ and\ \bibinfo {author} {\bibfnamefont {J.~K.}\ \bibnamefont
  {Freericks}},\ }\href@noop {} {\bibfield  {journal} {\bibinfo  {journal}
  {Phys. Rev. A}\ }\textbf {\bibinfo {volume} {83}},\ \bibinfo {pages} {043609}
  (\bibinfo {year} {2011})}\BibitemShut {NoStop}%
\bibitem [{\citenamefont {Parish}\ and\ \citenamefont
  {Marchetti}(2012)}]{Parish2012}%
  \BibitemOpen
  \bibfield  {author} {\bibinfo {author} {\bibfnamefont {M.~M.}\ \bibnamefont
  {Parish}}\ and\ \bibinfo {author} {\bibfnamefont {F.~M.}\ \bibnamefont
  {Marchetti}},\ }\href@noop {} {\bibfield  {journal} {\bibinfo  {journal}
  {Phys. Rev. Lett.}\ }\textbf {\bibinfo {volume} {108}},\ \bibinfo {pages}
  {145304} (\bibinfo {year} {2012})}\BibitemShut {NoStop}%
\bibitem [{\citenamefont {Block}\ and\ \citenamefont
  {Bruun}(2014)}]{Block2014}%
  \BibitemOpen
  \bibfield  {author} {\bibinfo {author} {\bibfnamefont {J.~K.}\ \bibnamefont
  {Block}}\ and\ \bibinfo {author} {\bibfnamefont {G.~M.}\ \bibnamefont
  {Bruun}},\ }\href@noop {} {\bibfield  {journal} {\bibinfo  {journal} {Phys.
  Rev. B}\ }\textbf {\bibinfo {volume} {90}},\ \bibinfo {pages} {155102}
  (\bibinfo {year} {2014})}\BibitemShut {NoStop}%
\bibitem [{\citenamefont {Matveeva}\ and\ \citenamefont
  {Giorgini}(2012)}]{Matveeva2012}%
  \BibitemOpen
  \bibfield  {author} {\bibinfo {author} {\bibfnamefont {N.}~\bibnamefont
  {Matveeva}}\ and\ \bibinfo {author} {\bibfnamefont {S.}~\bibnamefont
  {Giorgini}},\ }\href@noop {} {\bibfield  {journal} {\bibinfo  {journal}
  {Phys. Rev. Lett.}\ }\textbf {\bibinfo {volume} {109}},\ \bibinfo {pages}
  {200401} (\bibinfo {year} {2012})}\BibitemShut {NoStop}%
\bibitem [{\citenamefont {Babadi}\ \emph {et~al.}(2013)\citenamefont {Babadi},
  \citenamefont {Skinner}, \citenamefont {Fogler},\ and\ \citenamefont
  {Demler}}]{Babadi2013}%
  \BibitemOpen
  \bibfield  {author} {\bibinfo {author} {\bibfnamefont {M.}~\bibnamefont
  {Babadi}}, \bibinfo {author} {\bibfnamefont {B.}~\bibnamefont {Skinner}},
  \bibinfo {author} {\bibfnamefont {M.~M.}\ \bibnamefont {Fogler}}, \ and\
  \bibinfo {author} {\bibfnamefont {E.}~\bibnamefont {Demler}},\ }\href@noop {}
  {\bibfield  {journal} {\bibinfo  {journal} {Europhys. Lett.}\ }\textbf
  {\bibinfo {volume} {103}},\ \bibinfo {pages} {16002} (\bibinfo {year}
  {2013})}\BibitemShut {NoStop}%
\bibitem [{\citenamefont {Baranov}\ \emph {et~al.}(2005)\citenamefont
  {Baranov}, \citenamefont {Osterloh},\ and\ \citenamefont
  {Lewenstein}}]{Baranov2005}%
  \BibitemOpen
  \bibfield  {author} {\bibinfo {author} {\bibfnamefont {M.~A.}\ \bibnamefont
  {Baranov}}, \bibinfo {author} {\bibfnamefont {K.}~\bibnamefont {Osterloh}}, \
  and\ \bibinfo {author} {\bibfnamefont {M.}~\bibnamefont {Lewenstein}},\
  }\href@noop {} {\bibfield  {journal} {\bibinfo  {journal} {Phys. Rev. Lett.}\
  }\textbf {\bibinfo {volume} {94}},\ \bibinfo {pages} {070404} (\bibinfo
  {year} {2005})}\BibitemShut {NoStop}%
\bibitem [{\citenamefont {Baranov}\ \emph {et~al.}(2008)\citenamefont
  {Baranov}, \citenamefont {Fehrmann},\ and\ \citenamefont
  {Lewenstein}}]{baranov2008wigner}%
  \BibitemOpen
  \bibfield  {author} {\bibinfo {author} {\bibfnamefont {M.~A.}\ \bibnamefont
  {Baranov}}, \bibinfo {author} {\bibfnamefont {H.}~\bibnamefont {Fehrmann}}, \
  and\ \bibinfo {author} {\bibfnamefont {M.}~\bibnamefont {Lewenstein}},\
  }\href@noop {} {\bibfield  {journal} {\bibinfo  {journal} {Phys. Rev. Lett.}\
  }\textbf {\bibinfo {volume} {100}},\ \bibinfo {pages} {200402} (\bibinfo
  {year} {2008})}\BibitemShut {NoStop}%
\bibitem [{\citenamefont {Zhang}\ and\ \citenamefont
  {Yi}(2010)}]{zhang2010thermodynamic}%
  \BibitemOpen
  \bibfield  {author} {\bibinfo {author} {\bibfnamefont {J.-N.}\ \bibnamefont
  {Zhang}}\ and\ \bibinfo {author} {\bibfnamefont {S.}~\bibnamefont {Yi}},\
  }\href@noop {} {\bibfield  {journal} {\bibinfo  {journal} {Phys. Rev. A}\
  }\textbf {\bibinfo {volume} {81}},\ \bibinfo {pages} {033617} (\bibinfo
  {year} {2010})}\BibitemShut {NoStop}%
\bibitem [{\citenamefont {Endo}\ \emph {et~al.}(2010)\citenamefont {Endo},
  \citenamefont {Miyakawa},\ and\ \citenamefont {Nikuni}}]{Endo2010}%
  \BibitemOpen
  \bibfield  {author} {\bibinfo {author} {\bibfnamefont {Y.}~\bibnamefont
  {Endo}}, \bibinfo {author} {\bibfnamefont {T.}~\bibnamefont {Miyakawa}}, \
  and\ \bibinfo {author} {\bibfnamefont {T.}~\bibnamefont {Nikuni}},\
  }\href@noop {} {\bibfield  {journal} {\bibinfo  {journal} {Phys. Rev. A}\
  }\textbf {\bibinfo {volume} {81}},\ \bibinfo {pages} {063624} (\bibinfo
  {year} {2010})}\BibitemShut {NoStop}%
\bibitem [{\citenamefont {Kestner}\ and\ \citenamefont
  {Das~Sarma}(2010)}]{Kestner2010}%
  \BibitemOpen
  \bibfield  {author} {\bibinfo {author} {\bibfnamefont {J.~P.}\ \bibnamefont
  {Kestner}}\ and\ \bibinfo {author} {\bibfnamefont {S.}~\bibnamefont
  {Das~Sarma}},\ }\href@noop {} {\bibfield  {journal} {\bibinfo  {journal}
  {Phys. Rev. A}\ }\textbf {\bibinfo {volume} {82}},\ \bibinfo {pages} {033608}
  (\bibinfo {year} {2010})}\BibitemShut {NoStop}%
\bibitem [{\citenamefont {Zhang}\ \emph {et~al.}(2011)\citenamefont {Zhang},
  \citenamefont {Qiu}, \citenamefont {He},\ and\ \citenamefont
  {Yi}}]{Zhang2011}%
  \BibitemOpen
  \bibfield  {author} {\bibinfo {author} {\bibfnamefont {J.-N.}\ \bibnamefont
  {Zhang}}, \bibinfo {author} {\bibfnamefont {R.-Z.}\ \bibnamefont {Qiu}},
  \bibinfo {author} {\bibfnamefont {L.}~\bibnamefont {He}}, \ and\ \bibinfo
  {author} {\bibfnamefont {S.}~\bibnamefont {Yi}},\ }\href@noop {} {\bibfield
  {journal} {\bibinfo  {journal} {Phys. Rev. A}\ }\textbf {\bibinfo {volume}
  {83}},\ \bibinfo {pages} {053628} (\bibinfo {year} {2011})}\BibitemShut
  {NoStop}%
\bibitem [{\citenamefont {G{\'o}ral}\ \emph {et~al.}(2003)\citenamefont
  {G{\'o}ral}, \citenamefont {Brewczyk},\ and\ \citenamefont
  {Rza̧{\.z}ewski}}]{Goral2003}%
  \BibitemOpen
  \bibfield  {author} {\bibinfo {author} {\bibfnamefont {K.}~\bibnamefont
  {G{\'o}ral}}, \bibinfo {author} {\bibfnamefont {M.}~\bibnamefont {Brewczyk}},
  \ and\ \bibinfo {author} {\bibfnamefont {K.}~\bibnamefont
  {Rza̧{\.z}ewski}},\ }\href@noop {} {\bibfield  {journal} {\bibinfo
  {journal} {Phys. Rev. A}\ }\textbf {\bibinfo {volume} {67}},\ \bibinfo
  {pages} {025601} (\bibinfo {year} {2003})}\BibitemShut {NoStop}%
\bibitem [{\citenamefont {Sogo}\ \emph {et~al.}(2009)\citenamefont {Sogo},
  \citenamefont {He}, \citenamefont {Miyakawa}, \citenamefont {Yi},
  \citenamefont {Lu},\ and\ \citenamefont {Pu}}]{Sogo2009}%
  \BibitemOpen
  \bibfield  {author} {\bibinfo {author} {\bibfnamefont {T.}~\bibnamefont
  {Sogo}}, \bibinfo {author} {\bibfnamefont {L.}~\bibnamefont {He}}, \bibinfo
  {author} {\bibfnamefont {T.}~\bibnamefont {Miyakawa}}, \bibinfo {author}
  {\bibfnamefont {S.}~\bibnamefont {Yi}}, \bibinfo {author} {\bibfnamefont
  {H.}~\bibnamefont {Lu}}, \ and\ \bibinfo {author} {\bibfnamefont
  {H.}~\bibnamefont {Pu}},\ }\href@noop {} {\bibfield  {journal} {\bibinfo
  {journal} {New J. Phys.}\ }\textbf {\bibinfo {volume} {11}},\ \bibinfo
  {pages} {055017} (\bibinfo {year} {2009})}\BibitemShut {NoStop}%
\bibitem [{\citenamefont {Lima}\ and\ \citenamefont
  {Pelster}(2010{\natexlab{a}})}]{Lima2010a}%
  \BibitemOpen
  \bibfield  {author} {\bibinfo {author} {\bibfnamefont {A.~R.~P.}\
  \bibnamefont {Lima}}\ and\ \bibinfo {author} {\bibfnamefont {A.}~\bibnamefont
  {Pelster}},\ }\href@noop {} {\bibfield  {journal} {\bibinfo  {journal} {Phys.
  Rev. A}\ }\textbf {\bibinfo {volume} {81}},\ \bibinfo {pages} {021606}
  (\bibinfo {year} {2010}{\natexlab{a}})}\BibitemShut {NoStop}%
\bibitem [{\citenamefont {Lima}\ and\ \citenamefont
  {Pelster}(2010{\natexlab{b}})}]{Lima2010}%
  \BibitemOpen
  \bibfield  {author} {\bibinfo {author} {\bibfnamefont {A.~R.~P.}\
  \bibnamefont {Lima}}\ and\ \bibinfo {author} {\bibfnamefont {A.}~\bibnamefont
  {Pelster}},\ }\href@noop {} {\bibfield  {journal} {\bibinfo  {journal} {Phys.
  Rev. A}\ }\textbf {\bibinfo {volume} {81}},\ \bibinfo {pages} {063629}
  (\bibinfo {year} {2010}{\natexlab{b}})}\BibitemShut {NoStop}%
\bibitem [{\citenamefont {Abad}\ \emph {et~al.}(2012)\citenamefont {Abad},
  \citenamefont {Recati},\ and\ \citenamefont {Stringari}}]{Abad2012}%
  \BibitemOpen
  \bibfield  {author} {\bibinfo {author} {\bibfnamefont {M.}~\bibnamefont
  {Abad}}, \bibinfo {author} {\bibfnamefont {A.}~\bibnamefont {Recati}}, \ and\
  \bibinfo {author} {\bibfnamefont {S.}~\bibnamefont {Stringari}},\ }\href@noop
  {} {\bibfield  {journal} {\bibinfo  {journal} {Phys. Rev. A}\ }\textbf
  {\bibinfo {volume} {85}},\ \bibinfo {pages} {033639} (\bibinfo {year}
  {2012})}\BibitemShut {NoStop}%
\bibitem [{\citenamefont {Babadi}\ and\ \citenamefont
  {Demler}(2012)}]{Babadi2012}%
  \BibitemOpen
  \bibfield  {author} {\bibinfo {author} {\bibfnamefont {M.}~\bibnamefont
  {Babadi}}\ and\ \bibinfo {author} {\bibfnamefont {E.}~\bibnamefont
  {Demler}},\ }\href@noop {} {\bibfield  {journal} {\bibinfo  {journal} {Phys.
  Rev. A}\ }\textbf {\bibinfo {volume} {86}},\ \bibinfo {pages} {063638}
  (\bibinfo {year} {2012})}\BibitemShut {NoStop}%
\bibitem [{\citenamefont {W{\"a}chtler}\ \emph {et~al.}(2013)\citenamefont
  {W{\"a}chtler}, \citenamefont {Lima},\ and\ \citenamefont
  {Pelster}}]{Wachtler2013}%
  \BibitemOpen
  \bibfield  {author} {\bibinfo {author} {\bibfnamefont {F.}~\bibnamefont
  {W{\"a}chtler}}, \bibinfo {author} {\bibfnamefont {A.~R.~P.}\ \bibnamefont
  {Lima}}, \ and\ \bibinfo {author} {\bibfnamefont {A.}~\bibnamefont
  {Pelster}},\ }\href@noop {} {\bibfield  {journal} {\bibinfo  {journal} {arXiv
  preprint arXiv:1311.5100}\ } (\bibinfo {year} {2013})}\BibitemShut {NoStop}%
\bibitem [{\citenamefont {Pikovski}\ \emph {et~al.}(2010)\citenamefont
  {Pikovski}, \citenamefont {Klawunn}, \citenamefont {Shlyapnikov},\ and\
  \citenamefont {Santos}}]{Pikovski2010}%
  \BibitemOpen
  \bibfield  {author} {\bibinfo {author} {\bibfnamefont {A.}~\bibnamefont
  {Pikovski}}, \bibinfo {author} {\bibfnamefont {M.}~\bibnamefont {Klawunn}},
  \bibinfo {author} {\bibfnamefont {G.~V.}\ \bibnamefont {Shlyapnikov}}, \ and\
  \bibinfo {author} {\bibfnamefont {L.}~\bibnamefont {Santos}},\ }\href@noop {}
  {\bibfield  {journal} {\bibinfo  {journal} {Phys. Rev. Lett.}\ }\textbf
  {\bibinfo {volume} {105}},\ \bibinfo {pages} {215302} (\bibinfo {year}
  {2010})}\BibitemShut {NoStop}%
\bibitem [{\citenamefont {Baranov}\ \emph {et~al.}(2011)\citenamefont
  {Baranov}, \citenamefont {Micheli}, \citenamefont {Ronen},\ and\
  \citenamefont {Zoller}}]{baranov2011bilayer}%
  \BibitemOpen
  \bibfield  {author} {\bibinfo {author} {\bibfnamefont {M.~A.}\ \bibnamefont
  {Baranov}}, \bibinfo {author} {\bibfnamefont {A.}~\bibnamefont {Micheli}},
  \bibinfo {author} {\bibfnamefont {S.}~\bibnamefont {Ronen}}, \ and\ \bibinfo
  {author} {\bibfnamefont {P.}~\bibnamefont {Zoller}},\ }\href@noop {}
  {\bibfield  {journal} {\bibinfo  {journal} {Phys. Rev. A}\ }\textbf {\bibinfo
  {volume} {83}},\ \bibinfo {pages} {043602} (\bibinfo {year}
  {2011})}\BibitemShut {NoStop}%
\bibitem [{\citenamefont {Zinner}\ \emph {et~al.}(2012)\citenamefont {Zinner},
  \citenamefont {Wunsch}, \citenamefont {Pekker},\ and\ \citenamefont
  {Wang}}]{Zinner2012}%
  \BibitemOpen
  \bibfield  {author} {\bibinfo {author} {\bibfnamefont {N.~T.}\ \bibnamefont
  {Zinner}}, \bibinfo {author} {\bibfnamefont {B.}~\bibnamefont {Wunsch}},
  \bibinfo {author} {\bibfnamefont {D.}~\bibnamefont {Pekker}}, \ and\ \bibinfo
  {author} {\bibfnamefont {D.-W.}\ \bibnamefont {Wang}},\ }\href@noop {}
  {\bibfield  {journal} {\bibinfo  {journal} {Phys. Rev. A}\ }\textbf {\bibinfo
  {volume} {85}},\ \bibinfo {pages} {013603} (\bibinfo {year}
  {2012})}\BibitemShut {NoStop}%
\bibitem [{\citenamefont {Matveeva}\ and\ \citenamefont
  {Giorgini}(2014)}]{Matveeva2014}%
  \BibitemOpen
  \bibfield  {author} {\bibinfo {author} {\bibfnamefont {N.}~\bibnamefont
  {Matveeva}}\ and\ \bibinfo {author} {\bibfnamefont {S.}~\bibnamefont
  {Giorgini}},\ }\href@noop {} {\bibfield  {journal} {\bibinfo  {journal}
  {Phys. Rev. A}\ }\textbf {\bibinfo {volume} {90}},\ \bibinfo {pages} {053620}
  (\bibinfo {year} {2014})}\BibitemShut {NoStop}%
\bibitem [{\citenamefont {Nessi}\ \emph {et~al.}(2014)\citenamefont {Nessi},
  \citenamefont {Iucci},\ and\ \citenamefont {Cazalilla}}]{Nessi2014}%
  \BibitemOpen
  \bibfield  {author} {\bibinfo {author} {\bibfnamefont {N.}~\bibnamefont
  {Nessi}}, \bibinfo {author} {\bibfnamefont {A.}~\bibnamefont {Iucci}}, \ and\
  \bibinfo {author} {\bibfnamefont {M.~A.}\ \bibnamefont {Cazalilla}},\
  }\href@noop {} {\bibfield  {journal} {\bibinfo  {journal} {Phys. Rev. Lett.}\
  }\textbf {\bibinfo {volume} {113}},\ \bibinfo {pages} {210402} (\bibinfo
  {year} {2014})}\BibitemShut {NoStop}%
\bibitem [{\citenamefont {Lu}\ and\ \citenamefont
  {Shlyapnikov}(2012)}]{Lu2012a}%
  \BibitemOpen
  \bibfield  {author} {\bibinfo {author} {\bibfnamefont {Z.-K.}\ \bibnamefont
  {Lu}}\ and\ \bibinfo {author} {\bibfnamefont {G.~V.}\ \bibnamefont
  {Shlyapnikov}},\ }\href@noop {} {\bibfield  {journal} {\bibinfo  {journal}
  {Phys. Rev. A}\ }\textbf {\bibinfo {volume} {85}},\ \bibinfo {pages} {023614}
  (\bibinfo {year} {2012})}\BibitemShut {NoStop}%
\bibitem [{\citenamefont {Liu}\ and\ \citenamefont {Yin}(2011)}]{Liu2011a}%
  \BibitemOpen
  \bibfield  {author} {\bibinfo {author} {\bibfnamefont {B.}~\bibnamefont
  {Liu}}\ and\ \bibinfo {author} {\bibfnamefont {L.}~\bibnamefont {Yin}},\
  }\href@noop {} {\bibfield  {journal} {\bibinfo  {journal} {Phys. Rev. A}\
  }\textbf {\bibinfo {volume} {84}},\ \bibinfo {pages} {053603} (\bibinfo
  {year} {2011})}\BibitemShut {NoStop}%
\bibitem [{GSL()}]{GSL}%
  \BibitemOpen
  \href@noop {} {\emph {\bibinfo {title} {GNU Scientific Library (GSL)}}},\
  \bibinfo {note}
  {\url{http://www.gnu.org/software/gsl/manual/html_node/VEGAS.html}}\BibitemShut
  {NoStop}%
\bibitem [{\citenamefont {Luttinger}(1960)}]{Luttinger1960a}%
  \BibitemOpen
  \bibfield  {author} {\bibinfo {author} {\bibfnamefont {J.~M.}\ \bibnamefont
  {Luttinger}},\ }\href@noop {} {\bibfield  {journal} {\bibinfo  {journal}
  {Phys. Rev.}\ }\textbf {\bibinfo {volume} {119}},\ \bibinfo {pages} {1153}
  (\bibinfo {year} {1960})}\BibitemShut {NoStop}%
\bibitem [{\citenamefont {Yamaji}\ \emph {et~al.}(2006)\citenamefont {Yamaji},
  \citenamefont {Misawa},\ and\ \citenamefont {Imada}}]{Yamaji2006}%
  \BibitemOpen
  \bibfield  {author} {\bibinfo {author} {\bibfnamefont {Y.}~\bibnamefont
  {Yamaji}}, \bibinfo {author} {\bibfnamefont {T.}~\bibnamefont {Misawa}}, \
  and\ \bibinfo {author} {\bibfnamefont {M.}~\bibnamefont {Imada}},\
  }\href@noop {} {\bibfield  {journal} {\bibinfo  {journal} {J. Phys. Soc.
  Jpn.}\ }\textbf {\bibinfo {volume} {75}},\ \bibinfo {pages} {094719}
  (\bibinfo {year} {2006})}\BibitemShut {NoStop}%
\bibitem [{\citenamefont {Carr}\ \emph {et~al.}(2010)\citenamefont {Carr},
  \citenamefont {Quintanilla},\ and\ \citenamefont {Betouras}}]{Carr2010}%
  \BibitemOpen
  \bibfield  {author} {\bibinfo {author} {\bibfnamefont {S.~T.}\ \bibnamefont
  {Carr}}, \bibinfo {author} {\bibfnamefont {J.}~\bibnamefont {Quintanilla}}, \
  and\ \bibinfo {author} {\bibfnamefont {J.~J.}\ \bibnamefont {Betouras}},\
  }\href@noop {} {\bibfield  {journal} {\bibinfo  {journal} {Phys. Rev. B}\
  }\textbf {\bibinfo {volume} {82}},\ \bibinfo {pages} {045110} (\bibinfo
  {year} {2010})}\BibitemShut {NoStop}%
\bibitem [{\citenamefont {Slizovskiy}\ \emph {et~al.}(2014)\citenamefont
  {Slizovskiy}, \citenamefont {Betouras}, \citenamefont {Carr},\ and\
  \citenamefont {Quintanilla}}]{Slizovskiy2014}%
  \BibitemOpen
  \bibfield  {author} {\bibinfo {author} {\bibfnamefont {S.}~\bibnamefont
  {Slizovskiy}}, \bibinfo {author} {\bibfnamefont {J.~J.}\ \bibnamefont
  {Betouras}}, \bibinfo {author} {\bibfnamefont {S.~T.}\ \bibnamefont {Carr}},
  \ and\ \bibinfo {author} {\bibfnamefont {J.}~\bibnamefont {Quintanilla}},\
  }\href@noop {} {\bibfield  {journal} {\bibinfo  {journal} {Phys. Rev. B}\
  }\textbf {\bibinfo {volume} {90}},\ \bibinfo {pages} {165110} (\bibinfo
  {year} {2014})}\BibitemShut {NoStop}%
\end{thebibliography}%

\end{document}